\documentclass[11pt,a4paper]{article}
\pdfoutput=1
\usepackage{jheppub}

\usepackage{url}
\usepackage{bm}

\allowdisplaybreaks[2]

\def\cN{\mathcal{N}}
\def\cO{\mathcal{O}}
\def\cP{\mathcal{P}}

\def\cR{\mathcal{R}}
\def\cS{\mathcal{S}}

\def\mint{\int_{-\infty}^\infty\!\cdots\!\int_{-\infty}^\infty}

\newcommand{\be}{\begin{equation}}
\newcommand{\ee}{\end{equation}}
\newcommand{\ba}{\begin{aligned}}
\newcommand{\ea}{\end{aligned}}

\def\({\left(}
\def\){\right)}

\newcommand{\pd}{\partial}

\def\vvec#1#2{\begin{pmatrix} #1 \\ #2 \end{pmatrix}}

\def\mat#1#2#3#4{\begin{pmatrix} #1 & #2 \\ #3 & #4  \end{pmatrix}}

\preprint{RUP-17-24}

\title{Perturbative/nonperturbative aspects of Bloch electrons in a honeycomb lattice}

\author{Yasuyuki Hatsuda}

\affiliation{Department of Physics, Rikkyo University,\\Toshima, Tokyo 171-8501, Japan}

\emailAdd{yhatsuda@rikkyo.ac.jp}

\abstract{
We revisit the spectral problem for Bloch electrons in a two-dimensional bipartite honeycomb lattice under a uniform magnetic field.
It is well-known that such a honeycomb structure is realized in graphene.
We present a systematic framework to compute the perturbative magnetic flux
expansions near two distinct band edges. We then analyze the nonperturbative bandwidth of the spectrum. It turns out that there is a novel similarity between the spectrum near the Dirac point in the honeycomb lattice and the spectrum in the supersymmetric sine-Gordon quantum mechanics. We finally confirm a nontrivial vacuum-instanton-bion threesome relationship. Our analysis heavily relies on numerical experiments.
}

\begin{document}

\maketitle

\renewcommand{\thefootnote}{\arabic{footnote}}
\setcounter{footnote}{0}
\setcounter{section}{0}

\section{Introduction}
In his seminal paper \cite{Hof}, Hofstadter considered a simple  two-dimensional lattice model for Bloch electrons in the presence of a uniform magnetic field. 
In spite of the simple physical setup, its electron spectrum turned out to be remarkably rich.
Now, the Hofstadter model is recognized to be significant both in theoretical physics and in experimental physics
(and also in mathematical physics).
Recently, a new aspect of this model was revealed in the context of topological string theory
on toric Calabi-Yau threefolds \cite{HKT}, based on \cite{GHM1} that is far from condensed matter physics originally. 
In \cite{HSX}, the correspondence was generalized to the triangular lattice system \cite{CW} and another Calabi-Yau geometry.

Motivated by these developments,
we here revisit a more interesting model, Bloch electrons on a hexagonal honeycomb lattice. 
As is well-known, graphene has such a honeycomb structure, and thus the honeycomb lattice
is one of the most exciting 2d electron systems.
A characteristic property of the honeycomb lattice is that it has a zero energy gap.
Around the zero-gap energy, the dispersion behaves linearly, and it forms the Dirac cone,
where electrons are effectively described by massless Dirac fermions.
Such singular points (referred to as Dirac points below) make graphene a quite rich matter.
The spectrum of Bloch electrons in the honeycomb lattice under a magnetic field was studied by Rammal in \cite{Rammal}.

In this paper, we focus on perturbative and nonperturbative aspects of the honeycomb lattice.
A relation to quantum geometry in a toric Calabi-Yau threefold will be studied in detail in \cite{ToAppear1}.
We here analyze the spectrum in the weak magnetic flux regime. It is well-known that in this regime, the spectrum is characterized by Landau levels.
In the case of the honeycomb lattice, there are two energy regions in which we can use perturbation theory.
One is around the Dirac point, $E=0$. This is the bottom (or top) of the upper (or lower) band corresponding to the zero-gap in the zero magnetic field limit. 
We can also consider the perturbation around the other band edge at the top. In our convention, this corresponds to $E=3$ (or $E=-3$ equivalently).
By turning on the magnetic field, the spectrum is approximated by the perturbative expansion of the (rescaled) magnetic flux $\phi:=2\pi \Phi/\Phi_0$, where $\Phi_0=hc/e$ is the magnetic flux unit and $\Phi$ is the magnetic flux per unit cell.
We find that the perturbative expansion around the top $E=3$ is given by
\be
E_\text{top}^\text{pert}(n,\phi)=3-\frac{2n+1}{2\sqrt{3}}\phi+\frac{2n^2+2n+1}{72}\phi^2+\cO(\phi^3).
\label{eq:E-top-pert}
\ee
On the other hand, the perturbative expansion near the Dirac point is more involved.
We find the following expansion:
\be
E_\text{Dirac}^\text{pert}(n,\phi)=\pm 3^{1/4}\sqrt{n \phi}\(1-\frac{n}{4\sqrt{3}}\phi-\frac{17n^2+16}{864}\phi^2+\cO(\phi^3) \).
\label{eq:E-cone-pert}
\ee
In these equations, the non-negative integer $n$ labels the Landau level of the spectrum.
Note that the singular behavior $E \sim \pm 3^{1/4}\sqrt{n \phi}$ at the Dirac point was first found by McClure long time ago \cite{McClure}.
We apply an idea by Bender and Wu in \cite{BW} to the Bloch electron system, 
and compute these perturbative expansions systematically. 
See also \cite{RB} on the similar approach.
Such an efficient way is helpful to study nonperturbative physics in the sprit of ``resurgence theory''.

One notices that the lowest Landau level $n=0$ at the Dirac point is very special.
Its perturbative expansion seems vanishing,
\be
E_\text{Dirac}^\text{pert}(0,\phi)=0,
\ee
to all orders in perturbation theory. 
In fact, one can check that $E=0$ is always one of band edges for any rational $\Phi/\Phi_0$.
The band edge $E=0$ does not receive any ``quantum corrections'' in $\phi$. It is protected due to a certain symmetry like supersymmetry.%
\footnote{If considering the anisotropic honeycomb lattice, this symmetry is broken, and the zero-gap
disappears~\cite{ESKH}.}

We are interested in nonperturbative corrections to the perturbative expansions above.
A good quantity to explore them is the bandwidth of the spectrum
since the bandwidth is purely nonperturbative in $\phi$, related to quantum tunneling effects.
Based on the numerical analysis, we observe that the bandwidth for $\phi=2\pi/Q$ with integer $Q$ actually scales as
\be
\ba
\Delta E_\text{top}^\text{band}(n,2\pi/Q) \sim \cO(e^{-\frac{A}{2\pi}Q}), \quad
\Delta E_\text{Dirac}^\text{band}(n,2\pi/Q) \sim \cO(e^{-\frac{A}{10\pi}Q}), \quad
Q \to \infty,
\ea
\ee 
where the constant $A=10.149\cdots$ is exactly given by an integral form \eqref{eq:A}.
Interestingly, we find that the spectrum near the Dirac point is very similar to that in the supersymmetric sine-Gordon quantum mechanics \cite{DU-zero, KSTU}.%
\footnote{More precisely, $E^2$ is similar to the energy in the SUSY sine-Gordon.}
This suggests that ``Cheshire Cat Resurgence'' in \cite{KSTU} is useful in exploring more details on the nonperturbative structure in the weak flux regime.

Finally, we check the following threesome relation among the fluctuations around the vacuum, one-instanton and bion (i.e., instanton--anti-instanton) saddles:
\be
\frac{\cP_\text{fluc}^\text{bion}}{\(\cP_\text{fluc}^\text{inst}\)^2}=\(\frac{\pd \cP_\text{fluc}^\text{vac}}{\pd n}\)^{-1},
\label{eq:NP-NP}
\ee
where $n$ is a quantum number of the spectrum, and these fluctuations are normalized appropriately.
This is naturally expected by the so-called ``perturbative/nonperturbative relation,'' originally found in quantum mechanics \cite{AC1, AC2, AHS, Alvarez}.
We would like to emphasize that the relation \eqref{eq:NP-NP} holds \textit{universally} in a wide class of quantum mechanical systems
including 2d Bloch electron systems, whose resurgent property has not yet been understood well.%
\footnote{Note that in the 2d Bloch electron systems the perturbative/nonperturbative relationship in \cite{AC1, AC2, AHS, Alvarez} does not seem to work naively \cite{ToAppear}.
Nevertheless the relation \eqref{eq:NP-NP} is still working even in these cases.}

The organization of this paper is as follows.
In the next section, we start by reviewing Bloch electrons in a bipartite honeycomb lattice.
We explain how to compute the spectrum if the magnetic field is turned on.
Section~\ref{sec:pert} is a main part of this paper. 
We consider the weak magnetic flux limit, and present a systematic method to compute the weak magnetic flux expansions
near $E=3$ and $E=0$. We also discuss nonperturbative corrections in the bandwidth, based on numerical experiments.
We check a perturbative-instanton-bion triangle relation.
In section~\ref{sec:SSG}, we give comments on nonperturbative corrections in the supersymmetric sine-Gordon quantum mechanics.
In section~\ref{sec:conclusion}, we give final remarks.
In appendix~\ref{sec:num}, we explain our numerical technique used in this paper.

\section{Bloch electrons in a honeycomb lattice}\label{sec:honeycomb}
We will start with a review of an electron system in a two-dimensional honeycomb lattice.
We fix our convention by following \cite{Rammal}.

\subsection{No magnetic flux}
\begin{figure}[t]
\begin{center}
    \includegraphics[width=0.8\linewidth]{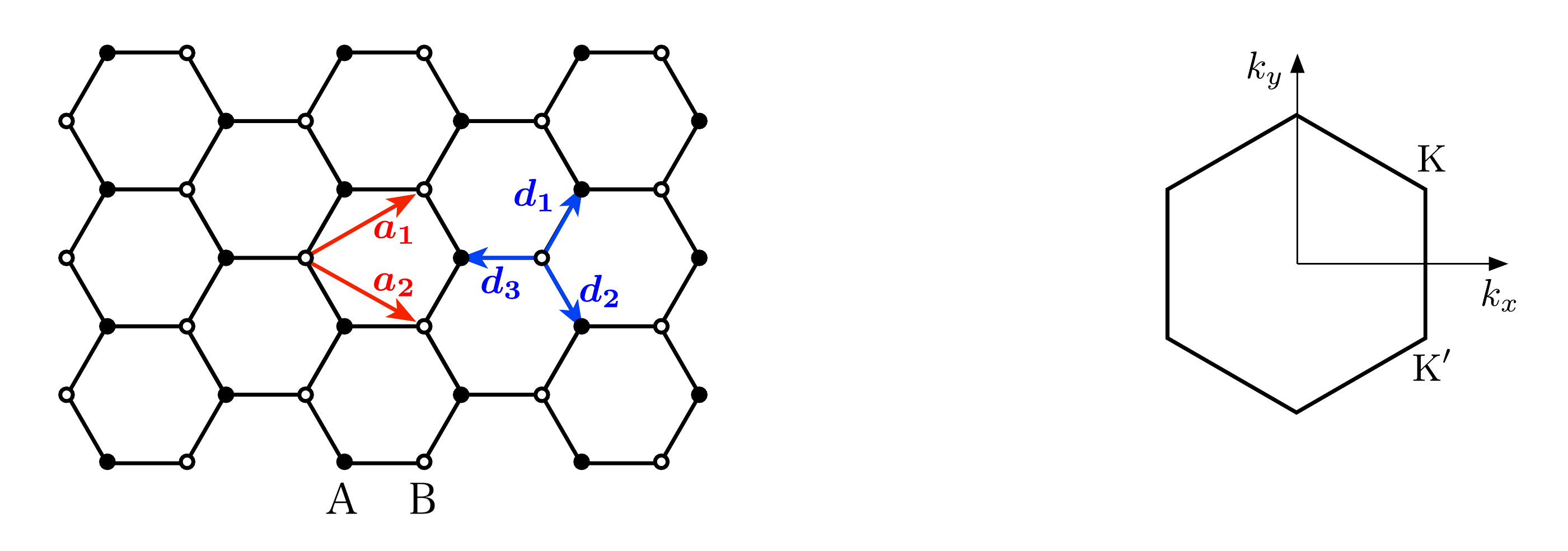}
\end{center}
  \caption{Left: The bipartite honeycomb lattice with lattice spacing $a$. Right: The first Brillouin zone in the reciprocal space.}
  \label{fig:honeycomb}
\end{figure}

In this subsection, we consider the case of no magnetic field.
The honeycomb lattice we are now treating is a bipartite system with two sublattices. We refer to them as A and B.
See the left of figure~\ref{fig:honeycomb}.
The primitive translation vectors are 
\be
\bm{a}_1=\biggl(\frac{3a}{2}, \frac{\sqrt{3}a}{2} \biggr),\qquad
\bm{a}_2=\biggl(\frac{3a}{2}, -\frac{\sqrt{3}a}{2} \biggr),
\label{eq:primitive-1}
\ee
where $a$ is a lattice spacing constant.
For later convenience, we also introduce three vectors in the real space, as shown in figure~\ref{fig:honeycomb} (left), by
\be
\bm{d}_1=\biggl( \frac{a}{2},\frac{\sqrt{3}a}{2} \biggr),\qquad
\bm{d}_2=\biggl( \frac{a}{2},-\frac{\sqrt{3}a}{2} \biggr),\qquad
\bm{d}_3=( -a, 0 ).\qquad
\label{eq:vec-d}
\ee
In the reciprocal space, the primitive vectors are given by
\be
\bm{b}_1=\biggl( \frac{2\pi}{3a}, \frac{2\pi}{\sqrt{3}a} \biggr),\qquad
\bm{b}_2=\biggl( \frac{2\pi}{3a}, -\frac{2\pi}{\sqrt{3}a} \biggr).
\label{eq:primitive-2}
\ee
The first Brillouin zone is a hexagon. Two corners of the hexagon are usually denoted by K and K$'$, whose wave vectors are given by
\be
\bm{K}=\biggl( \frac{2\pi}{3a}, \frac{2\pi}{3\sqrt{3}a} \biggr),\qquad
\bm{K}'=\biggl( \frac{2\pi}{3a}, -\frac{2\pi}{3\sqrt{3}a} \biggr).
\ee
See the right of figure \ref{fig:honeycomb}.

\begin{figure}[t]
\begin{center}
    \includegraphics[width=0.4\linewidth]{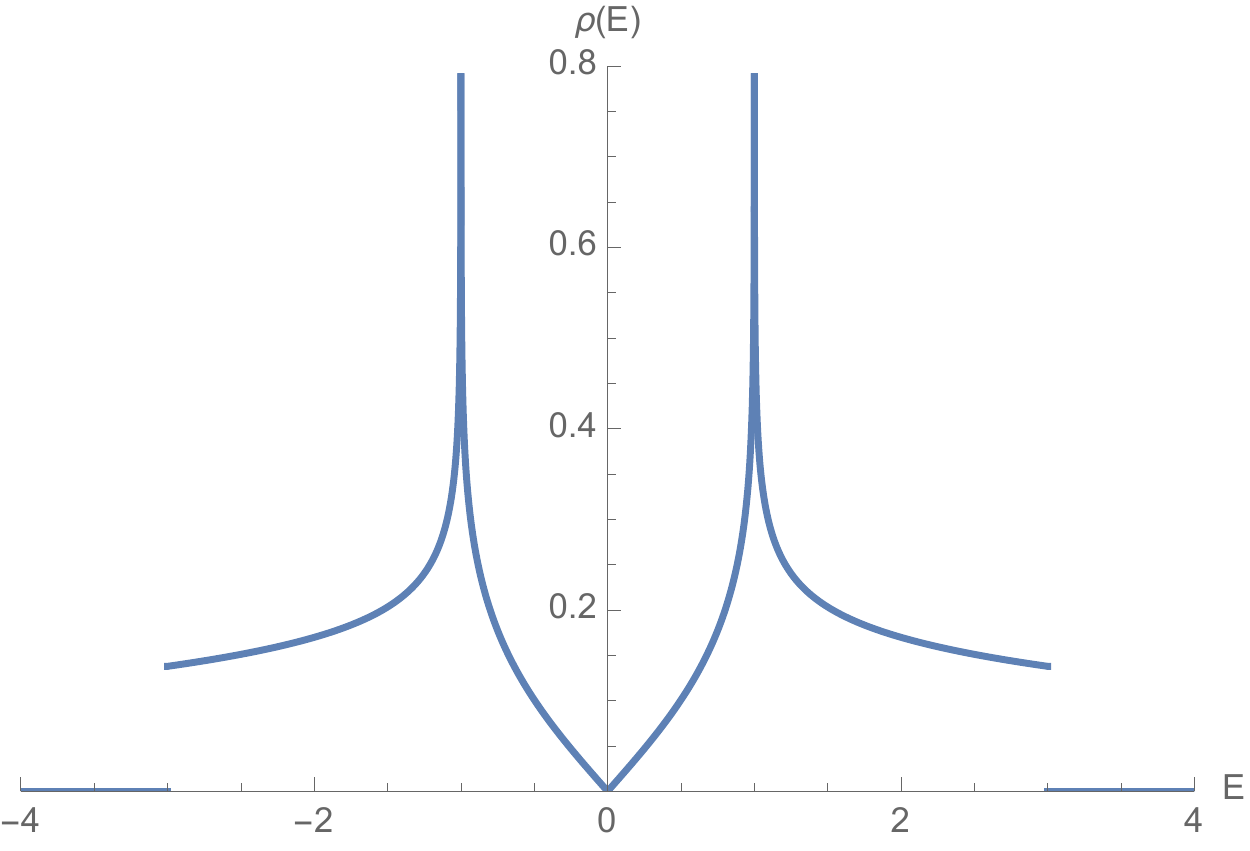}
\end{center}
  \caption{The density of states for the honeycomb lattice without magnetic field.}
  \label{fig:DOS}
\end{figure}

Let us proceed to the dispersion relation.
In the tight-binding approximation, the wave function with wave vector $\bm{k}$ is written as
\be
\psi_{\bm{k}}(\bm{r})=\sum_{X=\text{A}, \text{B}} \sum_{\bm{R}_l} e^{i \bm{k}\cdot \bm{R}_l} c_X(\bm{k}) \phi(\bm{r}-\bm{R}_l),
\ee
where $\bm{R}_\text{A}=n_1 \bm{a}_1+n_2 \bm{a}_2+\bm{d}_1$ and $\bm{R}_\text{B}=n_1 \bm{a}_1+n_2 \bm{a}_2$
($(n_1, n_2) \in \mathbb{Z}^2$) label positions of sublattices A and B respectively.
If considering only the nearest neighboring hopping, one obtains the following eigenvalue equations:
\be
\mat{0}{D(\bm{k})}{D(\bm{k})^*}{0} \vvec{c_\text{A}}{c_\text{B}}=E(\bm{k}) \vvec{c_\text{A}}{c_\text{B}},
\label{eq:eigen-eq1}
\ee
where $D(\bm{k}):=e^{-i\bm{k}\cdot \bm{d}_1}+e^{-i\bm{k}\cdot \bm{d}_2}+e^{-i \bm{k} \cdot \bm{d}_3}$.
Therefore the eigenvalues are given by $E(\bm{k})=\pm |D(\bm{k})|$.
Plugging \eqref{eq:vec-d}, we get the dispersion relation
\be
E(\bm{k})^2=3+2\cos(\sqrt{3}k_y a)+4\cos \biggl( \frac{3k_x a}{2} \biggr)
\cos \biggl( \frac{\sqrt{3}k_y a}{2} \biggr).
\ee
The range of the allowed energy is $-3 \leq E(\bm{k}) \leq 3$. 
Obviously, it is symmetric under $E \to -E$.
For a given wave vector $\bm{k}$, we have two eigenvalues $E(\bm{k})=\pm | D(\bm{k}) |$,
but at the points K and K$'$, these coincide, since $E(\bm{K})=E(\bm{K}')=0$.
Therefore these points are zero-gap points.
\begin{figure}[t]
\begin{center}
  \begin{minipage}[b]{0.4\linewidth}
    \centering
    \includegraphics[height=4cm]{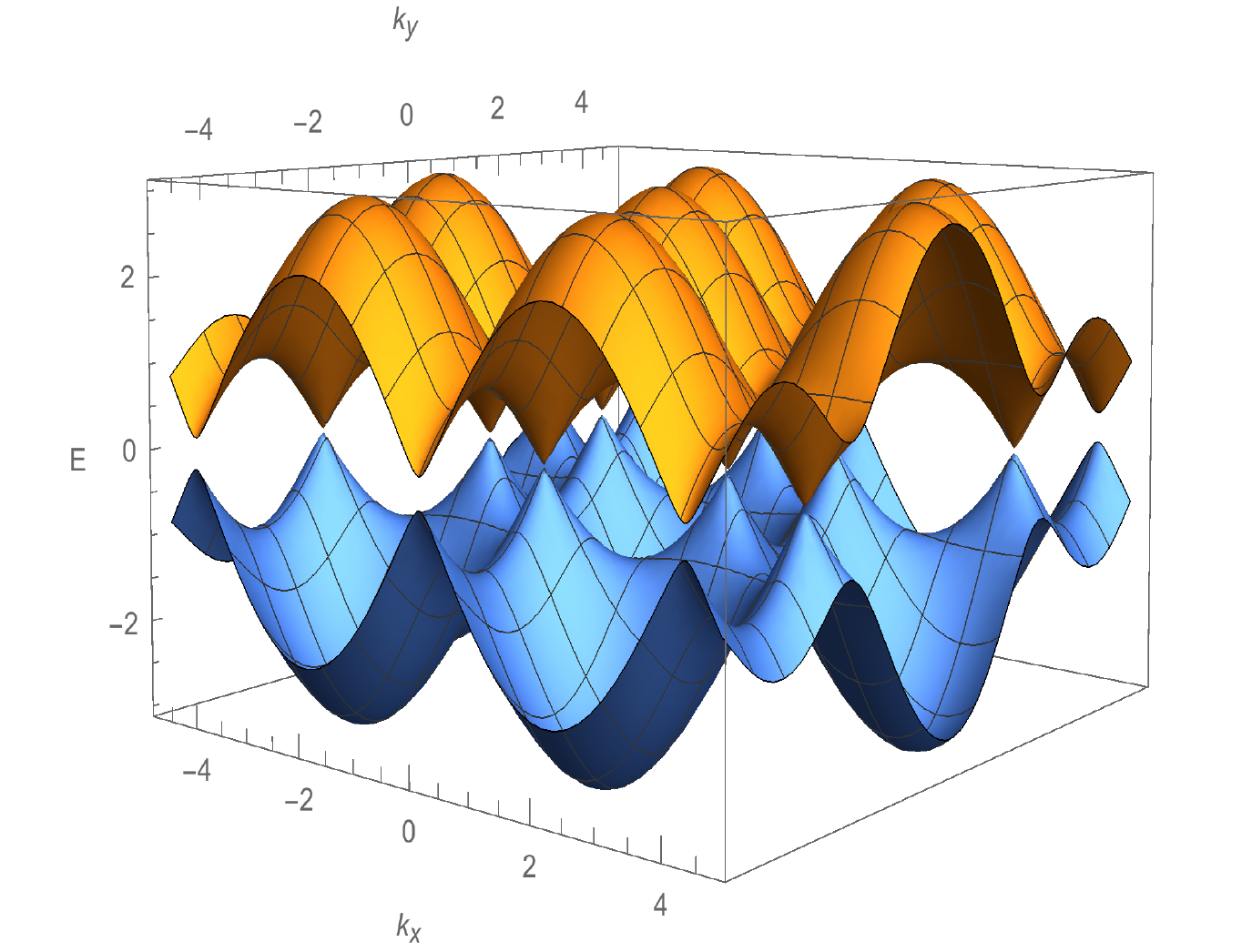}
  \end{minipage} \hspace{1cm}
  \begin{minipage}[b]{0.4\linewidth}
    \centering
    \includegraphics[height=4cm]{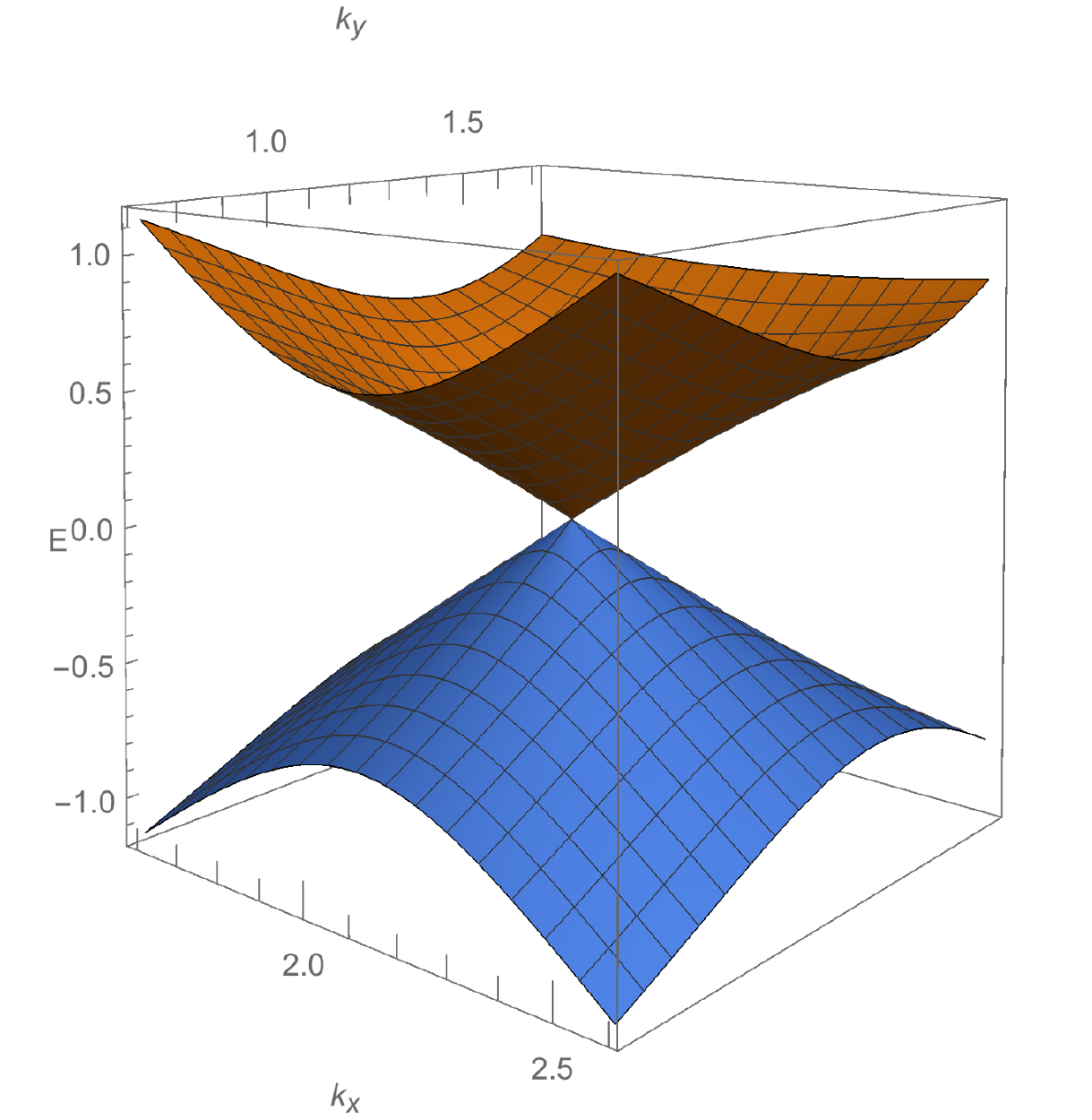}
  \end{minipage} 
\end{center}
  \caption{Left: The energy dispersion. Right: Zooming in a neighbor of a Dirac point.}
  \label{fig:dispersion}
\end{figure}
The dispersion is shown in figure~\ref{fig:dispersion} (left).
Since the dispersion near the zero-gap point K behaves as
\be
E(\bm{K}+\bm{k}) \approx \pm \frac{3a}{2}|\bm{k}|,
\ee
the energy surface forms two cones around this point, as in figure~\ref{fig:dispersion} (right). 
These are called the Dirac cones,
and we refer to the points K and K$'$ as the Dirac points, near which electrons are effectively described by relativistic massless fermions.
Finally, the density of states is given by
\be
\rho(E)=\begin{cases}
\displaystyle
\frac{\sqrt{|E|}}{2\pi^2 m^2} \mathbb{K}(1/m) \qquad &(0 \leq |E|  <1) ,\vspace{0.2cm}\\
\displaystyle
\frac{\sqrt{|E|}}{2\pi^2} \mathbb{K}(m) \qquad &(1<|E| \leq 3) ,\vspace{0.2cm}\\
0 \qquad &(|E|>3),
\end{cases}
\ee
where $m=(2|E|)^{-1}(1+|E|)^3(3-|E|)$ and $\mathbb{K}(m)$ is the complete elliptic integral of first kind. 
We plot $\rho(E)$ in figure~\ref{fig:DOS}. The density of states turns out to diverge at $E=\pm 1$ (a.k.a. Van Hove singularities).
Also it behaves as $\rho(E)=|E|/(\sqrt{3}\pi)+\cO(|E|^3)$ near $E=0$.

\subsection{Turning on magnetic flux}
If the magnetic field is turned on, the spectrum of the electron becomes richer and more involved.
In this case, one can effectively replace the wave vector by
\be
\hbar \bm{k} \to \bm{p}-\frac{e}{c}\bm{A}=-i\hbar \nabla -\frac{e}{c} \bm{A}.
\ee
Then the functions $D(\bm{k})$ and $D(\bm{k})^*$ are replaced by difference operators.
In the Landau gauge $\bm{A}=(0,Bx,0)$, these difference operators take the form of
\be
\ba
D(\bm{k}) \to \hat{D}&=e^{-\frac{i\phi}{12}}e^{\frac{i\phi}{3a}x} e^{-\frac{a}{2}\pd_x-\frac{\sqrt{3}a}{2}\pd_y}
+e^{\frac{i\phi}{12}}e^{-\frac{i\phi}{3a}x} e^{-\frac{a}{2}\pd_x+\frac{\sqrt{3}a}{2}\pd_y}+e^{a \pd_x}, \\
D(\bm{k})^* \to \hat{D}^\dagger&=e^{-\frac{i\phi}{12}}e^{-\frac{i\phi}{3a}x} e^{\frac{a}{2}\pd_x+\frac{\sqrt{3}a}{2}\pd_y}
+e^{\frac{i\phi}{12}}e^{\frac{i\phi}{3a}x} e^{\frac{a}{2}\pd_x-\frac{\sqrt{3}a}{2}\pd_y}+e^{-a \pd_x},
\ea
\ee
where the phase factors $e^{\pm i \phi/12}$ come from the Baker-Campbell-Hausdorff formula, and
\be
\phi:=2\pi \frac{\Phi}{\Phi_0},\qquad \Phi_0=\frac{hc}{e},\qquad \Phi=\frac{3\sqrt{3}a^2}{2}B.
\ee
Since the area of the hexagonal unit cell is $S_\text{unit cell}=3\sqrt{3}a^2/2$, $\Phi$ measures the magnetic flux per unit cell.
The eigenvalue equation \eqref{eq:eigen-eq1} is now replaced by the following two-dimensional difference equations:
\be
\ba
E \Psi_\text{A}(x,y)&=e^{\frac{i\phi}{3a}x-\frac{i\phi}{12}}\Psi_\text{B}\biggl( x-\frac{a}{2}, y-\frac{\sqrt{3}a}{2} \biggr)
+e^{-\frac{i\phi}{3a}x+\frac{i\phi}{12}}\Psi_\text{B}\biggl( x-\frac{a}{2}, y+\frac{\sqrt{3}a}{2} \biggr)
+\Psi_\text{B}(x+a,y), \\
E \Psi_\text{B}(x,y)&=e^{-\frac{i\phi}{3a}x-\frac{i\phi}{12}}\Psi_\text{A}\biggl( x+\frac{a}{2}, y+\frac{\sqrt{3}a}{2} \biggr)
+e^{\frac{i\phi}{3a}x+\frac{i\phi}{12}}\Psi_\text{A}\biggl( x+\frac{a}{2}, y-\frac{\sqrt{3}a}{2} \biggr)
+\Psi_\text{A}(x-a,y).
\ea
\ee
In the $y$-direction, we can take the plane wave solution by $\Psi_X(x,y)=e^{ik_y y}\psi_X(x)$, and the problem
reduces to the following one-dimensional problem:
\be
\ba
E \psi_\text{A}(x)&=2\cos\biggl( \frac{\phi}{3a}x-\frac{\phi}{12}-\frac{\sqrt{3}a}{2}k_y \biggr)
\psi_\text{B}\biggl( x-\frac{a}{2} \biggr)
+\psi_\text{B}(x+a), \\
E \psi_\text{B}(x)&=2\cos\biggl( \frac{\phi}{3a}x+\frac{\phi}{12}-\frac{\sqrt{3}a}{2}k_y \biggr)
\psi_\text{A}\biggl( x+\frac{a}{2} \biggr)
+\psi_\text{A}(x-a).
\ea 
\ee
By eliminating the unknown function $\psi_\text{A}(x)$, one gets the difference equation for $\psi_\text{B}(x)$:
\be
\ba
&\lambda \psi_\text{B}(x)=2\cos \biggl( \frac{\phi}{3a}x+\frac{\phi}{12}-\frac{\sqrt{3}a}{2}k_y\biggr) \psi_\text{B}\(x+\frac{3a}{2}\), \\
&+2\cos \biggl( \frac{\phi}{3a}x-\frac{5\phi}{12}-\frac{\sqrt{3}a}{2}k_y\biggr) \psi_\text{B}\(x-\frac{3a}{2}\)
+2\cos \biggl( \frac{2\phi}{3a}x+\frac{\phi}{6}-\sqrt{3}ak_y \biggr) \psi_\text{B}(x),
\ea
\label{eq:diff-1}
\ee
where $\lambda:=E^2-3$.
Let us write $\psi_m=\psi_\text{B}(\frac{3a}{2}m+\frac{a}{2})$.
Then the above eigenvalue equation is finally reduced to so-called Harper's equation:
\be
\ba
\lambda \psi_m=2\cos \( \frac{\phi}{2}m+\frac{\phi}{4}+\kappa\) \psi_{m+1}
+2\cos \( \frac{\phi}{2}m-\frac{\phi}{4}+\kappa\) \psi_{m-1}\\
+2\cos \(\phi m+\frac{\phi}{2}+2\kappa \) \psi_m,
\ea
\label{eq:Harper}
\ee
where $\kappa=-\sqrt{3}a k_y/2$.

\begin{figure}[t]
\begin{center}
  \begin{minipage}[b]{0.4\linewidth}
    \centering
    \includegraphics[height=6cm]{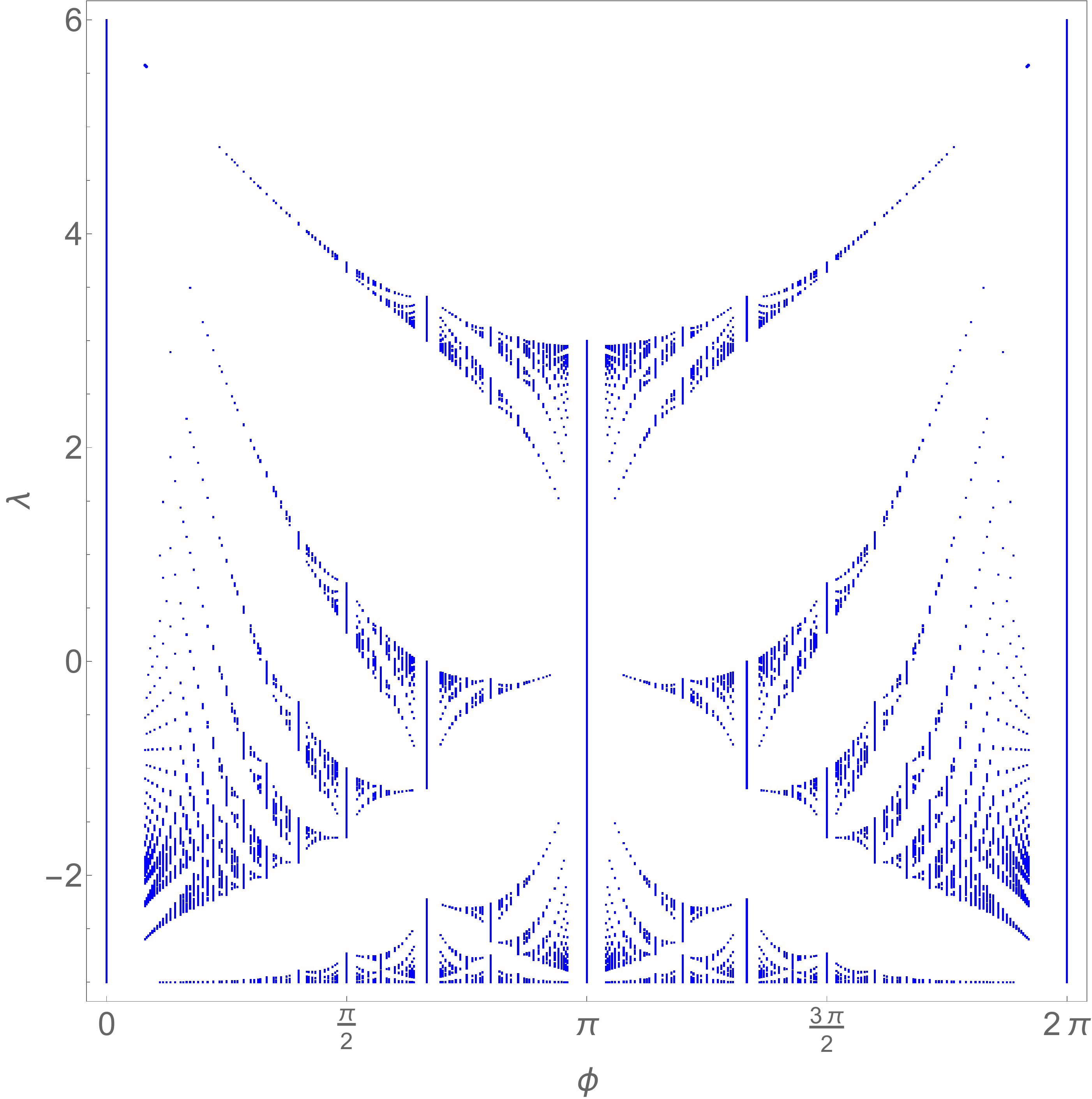}
  \end{minipage} \hspace{1cm}
  \begin{minipage}[b]{0.4\linewidth}
    \centering
    \includegraphics[height=6cm]{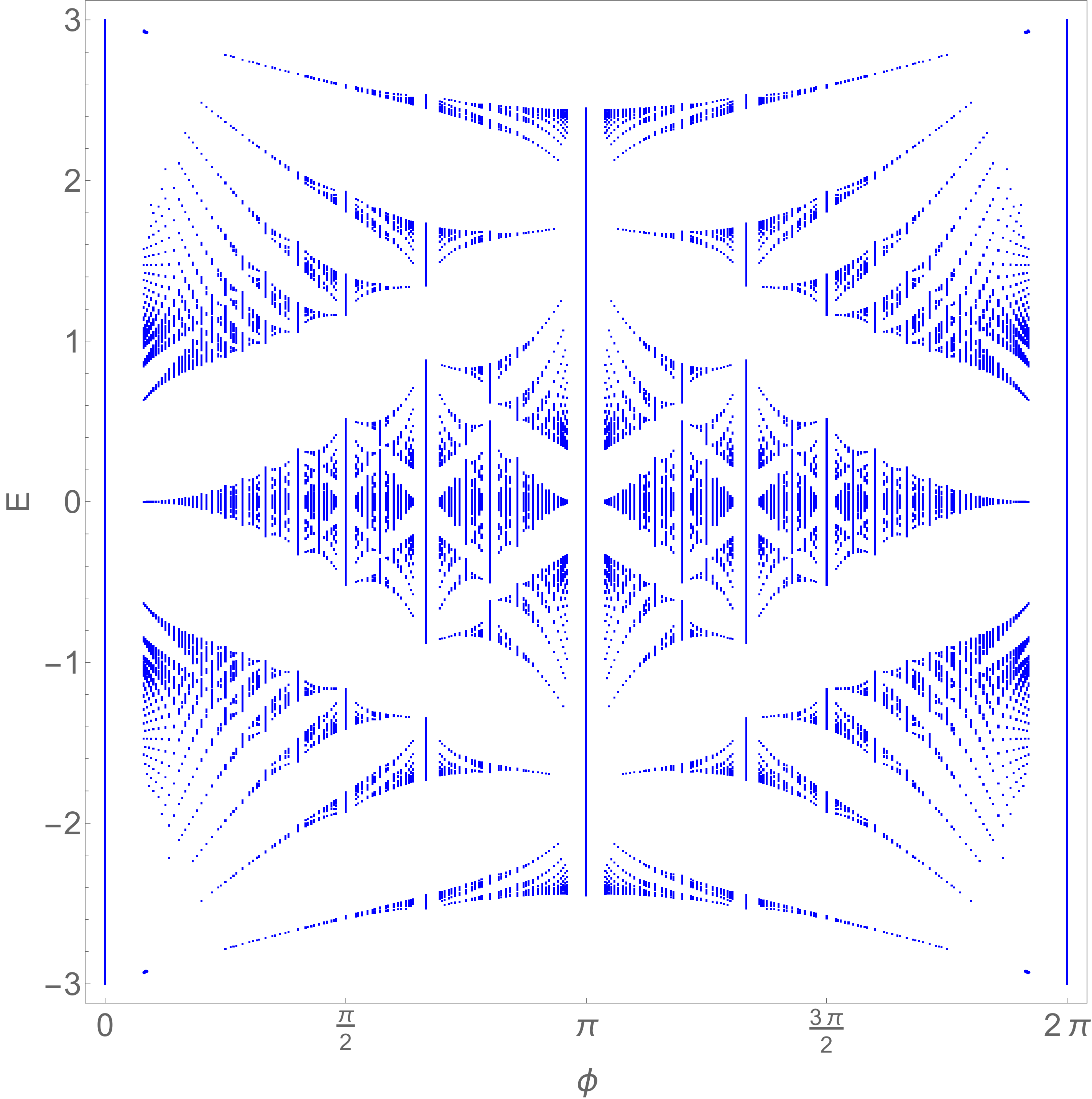}
  \end{minipage} 
\end{center}
  \caption{The spectrum of $\lambda$ (left) and of $E$ (right) as a function of the magnetic flux $\phi$. We plotted them for $\phi=2\pi a/b$ with all coprime integers $1\leq a < b \leq 25$.}
  \label{fig:butterfly}
\end{figure}

If $\phi/(2\pi)=\Phi/\Phi_0$ is rational, the eigenvalue problem of Harper's equation \eqref{eq:Harper}
is reduced to the diagonalization of a finite dimensional matrix $A$.
Let us assume $\phi=2\pi a/b $ with coprime integers $a$ and $b$.
As shown in \cite{Rammal}, the matrix elements of $A$ are explicitly given by
\be
\ba
A_{jj}&=2\cos\( \theta_1-\frac{2j+1}{2} \phi \), \qquad j=1,\dots,b, \\
A_{j,j+1}&=A_{j+1,j}^*=1+\exp i\(\theta_1-\frac{2j+1}{2} \phi \) ,\qquad j=1,\dots,b-1, \\
A_{b,1}&=A_{1,b}^*=e^{ib \theta_2} \left[ 1+\exp i\( \theta_1-\frac{2b+1}{2}\phi \) \right],
\ea
\ee
where $\theta_1=2\kappa$ and $\theta_2$ is Bloch's angle due to the periodicity of Harper's equation.
We can easily compute the eigenvalues of $A$ for given angles $\theta_1$ and $\theta_2$.
These form finite $b$ subbands of $\lambda$.
We show the spectra of $\lambda$ and $E=\pm \sqrt{\lambda+3}$ for $0 \leq \phi \leq 2\pi$ in figure~\ref{fig:butterfly}.%
\footnote{We thank Zhaojie Xu for drawing these figures.}

\section{Perturbative/nonperturbative aspects}\label{sec:pert}
In this section, we consider the weak magnetic flux limit $\phi \sim 0$ in the honeycomb system.
As is well-known, in the weak flux limit, the spectrum is characterized by Landau levels.
We first discuss the perturbative expansion in $\phi$, and then 
proceed to nonperturbative effects in the spectrum.

\subsection{Perturbative expansions}
To compute the perturbative expansion of the spectrum,
we first rewrite the eigenvalue equation \eqref{eq:diff-1} as a more symmetric form.
By shifting the argument 
\be
x \to x+\frac{a}{2}+\frac{3\sqrt{3}a^2}{2}k_y, \qquad
\psi_\text{B}\biggl( x+\frac{a}{2}+\frac{3\sqrt{3}a^2}{2}k_y \biggr) =: \psi(x),
\ee
we obtain
\be\ba
\lambda \psi(x)=2\cos\( \frac{\phi}{3a}x+\frac{\phi}{4} \)\psi\(x+\frac{3a}{2} \)
+2\cos\( \frac{\phi}{3a}x-\frac{\phi}{4} \) \psi\( x-\frac{3a}{2} \) \\
+2\cos \( \frac{2\phi}{3a}x+\frac{\phi}{2} \) \psi(x).
\ea\ee
The corresponding difference operator to the right hand side is
\be
H=2\cos\( \frac{\phi}{3a}x+\frac{\phi}{4} \) e^{\frac{3a}{2}\pd_x}
+2\cos\( \frac{\phi}{3a}x-\frac{\phi}{4} \)e^{-\frac{3a}{2}\pd_x}
+2\cos \( \frac{2\phi}{3a}x+\frac{\phi}{2} \).
\ee
After introducing a new variable $q:=\phi x/(3a)$, one obtains
\be
\ba
H=2\cos\( q+\frac{\phi}{4} \) e^{\frac{\phi}{2}\pd_q}
+2\cos\( q-\frac{\phi}{4} \)e^{-\frac{\phi}{2}\pd_q}
+2\cos \( 2q+\frac{\phi}{2} \).
\ea
\label{eq:H-1}
\ee
Recall that the eigenvalues of this operator gives $\lambda=E^2-3$ rather than $E$.

Let us first consider the zero flux limit.
The classical curve for \eqref{eq:H-1} is
\be
\lambda=4\cos q \cos \frac{p}{2}+2\cos 2q.
\label{eq:curve}
\ee
Therefore $\lambda$ takes the maximal value $\lambda=6$ at $p=q=0$ 
and the minimal value $\lambda=-3$ at $p=0, q=2\pi/3$.
Here $\lambda=6$ corresponds to the top of the band $E=3$, while $\lambda=-3$ to the zero-gap energy $E=0$.
Around these points, $H$ can be expanded as a perturbation of the harmonic oscillator.

We further rescale the variables by $q \to g q$ and $p \to g p$, where $g:=\sqrt{\phi}$.
We obtain
\be
H=2\cos\( gq+\frac{g^2}{4}\) e^{\frac{g}{2}\pd_q}
+2\cos\( gq-\frac{g^2}{4}\) e^{-\frac{g}{2}\pd_q}
+2\cos \( 2gq+\frac{g^2}{2} \).
\ee
Expanding it around $g=0$, we find
\be
H=6-\( -\frac{1}{2}\pd_q^2+6q^2\)g^2-2q g^{3}
+\frac{g^4}{96}(\pd_q^4-24q^2 \pd_q^2-48 q \pd_q+144q^4-36)+\cO(g^{5}).
\label{eq:H-pert}
\ee
This is the perturbative expansion around the top of the band.
The coefficient at order $g^2$ is just the Hamiltonian of the harmonic oscillator with $\hbar=m=1$
and $\omega=2\sqrt{3}$.
Therefore up to this order, the spectrum should be given by
\be
\lambda_\text{top}^\text{pert}=6-2\sqrt{3}\(n+\frac{1}{2}\)g^2+\cO(g^3).
\ee
where the quantum number $n$ represents the Landau level.
Since the Hamiltonian \eqref{eq:H-pert} can be regarded as the perturbation of the harmonic oscillator,
one can compute the higher order corrections by the textbook method.
To compute it more efficiently, the idea of Bender and Wu in \cite{BW} is extremely useful.\footnote{The similar analysis is found in \cite{RB}. However, our method here seems to be much efficient than theirs.}
According to \cite{BW}, the normalizable perturbative wave function of the operator \eqref{eq:H-pert} takes the form of
\be
\psi_n(q)=e^{-\sqrt{3}q^2} \sum_{\ell =0}^\infty g^\ell P_{n}^{(\ell)}(q),
\label{eq:BW}
\ee
where $P_n^{(\ell)}(q)$ is a polynomial of $q$.
For instance, the ground state wave function is given by
\be
\psi_0(q)=e^{-\sqrt{3}q^2}\biggl[ 1-\frac{q}{\sqrt{3}}g+\frac{q^2}{6} g^2+\( -\frac{q}{18}+\frac{q^3}{6\sqrt{3}} \) g^3+\cO(g^4) \biggr],
\ee
with the perturbative spectrum:
\be
\lambda_0^\text{pert}=6-\sqrt{3}g^2+\frac{g^4}{6}+\cO(g^6).
\ee
Assuming the Bender-Wu structure \eqref{eq:BW} to all orders in perturbation theory, one can determine $P_n^{(\ell)}(q)$ and the perturbative corrections to $\lambda$ recursively.
This enables us to put the computation on a computer. See \cite{SU-BW, GS-BW} for example.
Using this method, we find the perturbative expansion of $\lambda$:
\be
\ba
\lambda_\text{top}^\text{pert}(n,\phi)=6-\sqrt{3}(2n+1)\phi+\frac{3n^2+3n+1}{6}\phi^2
-\frac{n(2n^2+3n+1)}{36\sqrt{3}}\phi^3 \\
+\frac{3n^4+6n^3+9n^2+6n+4}{2592}\phi^4+\cO(\phi^5).
\ea
\label{eq:lambda-top-pert}
\ee
Plugging this expansion into $E=\pm \sqrt{\lambda+3}$, one obtains the expansion \eqref{eq:E-top-pert} of the electron energy.

\begin{figure}[t]
\begin{center}
  \begin{minipage}[b]{0.4\linewidth}
    \centering
    \includegraphics[height=5cm]{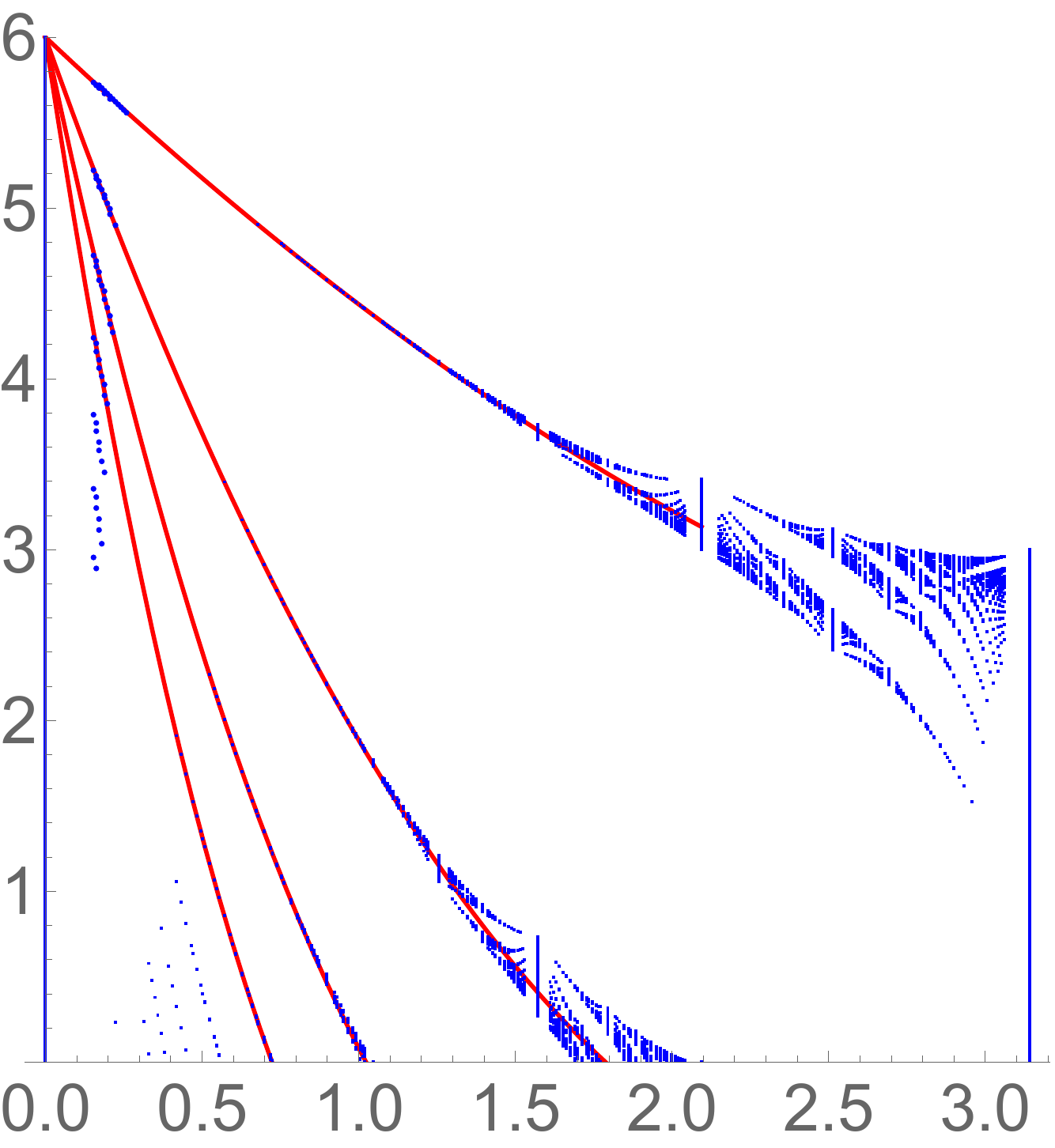}
  \end{minipage} \hspace{1cm}
  \begin{minipage}[b]{0.4\linewidth}
    \centering
    \includegraphics[height=5cm]{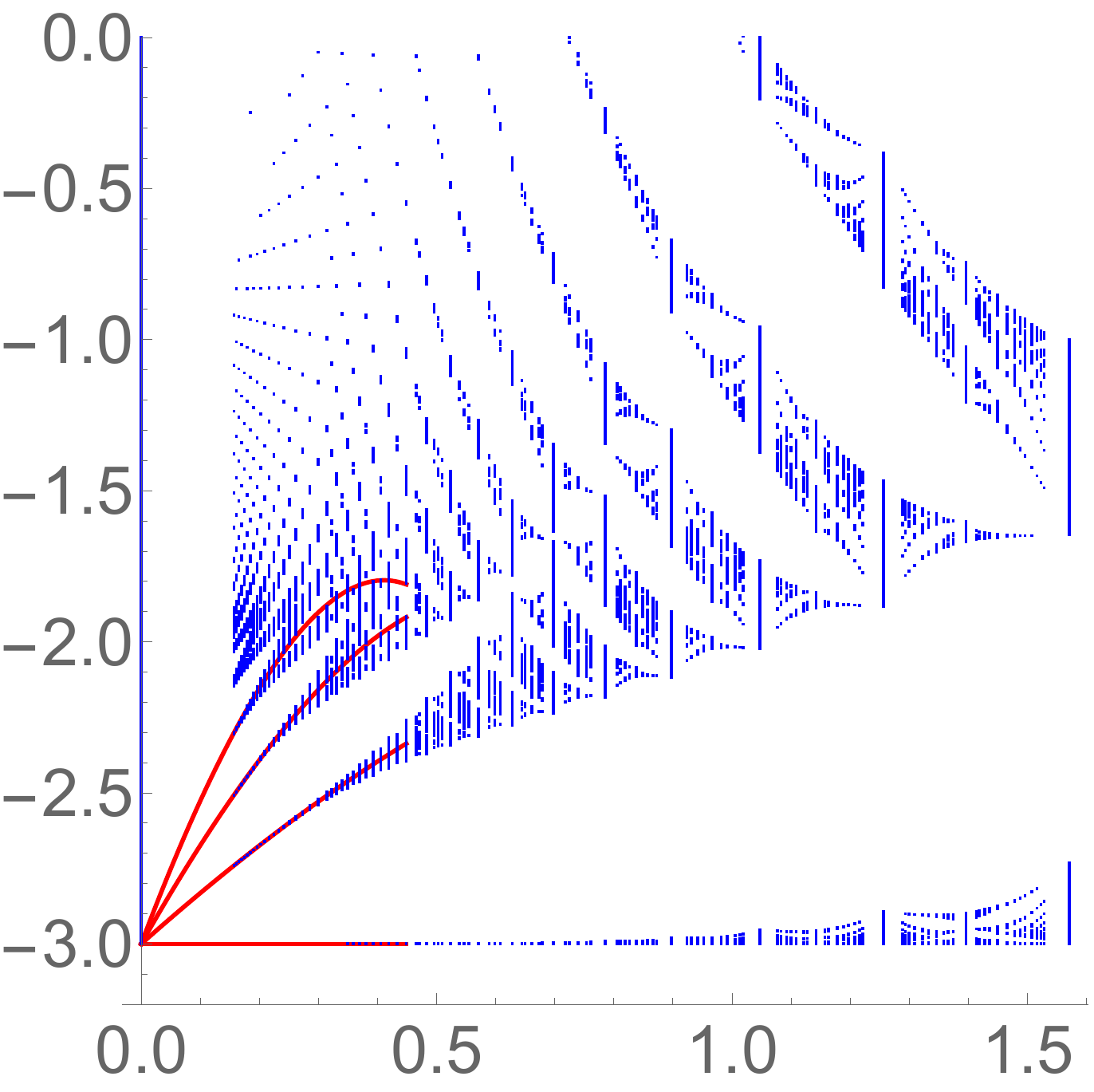}
  \end{minipage} 
\end{center}
  \caption{The perturbative expansions \eqref{eq:lambda-top-pert} and \eqref{eq:lambda-cone-pert} explain the Landau level splitting 
  in the weak magnetic limit. The red solid lines represent the perturbative expansions up to $\cO(\phi^4)$ for the first four Landau levels.}
  \label{fig:Landau}
\end{figure}

To compute the expansion around $\lambda=-3$, we first shift the position by $q \to q-2\pi/(3g)$,
and then expand the Hamiltonian around $g=0$. The result is
\be
H=-3+\(-\frac{\pd_q^2}{4}+3q^2-\frac{\sqrt{3}}{2} \) g^2+
\( \frac{\sqrt{3}q}{4}\pd_q^2+\frac{\sqrt{3}}{4}\pd_q+\sqrt{3}q^3+q\) g^3+\cO(g^4).
\ee 
In this case, we consider the perturbation of the harmonic oscillator with $\hbar=1$, $m=2$ and $\omega=\sqrt{3}$.
As in the same computation above, the perturbative expansion around the Dirac point results in
\be
\ba
&\lambda_\text{Dirac}^\text{pert}(n,\phi)=-3+\sqrt{3} n \phi-\frac{n^2}{2} \phi^2
-\frac{n(n^2+2)}{18\sqrt{3}}\phi^3-\frac{n^2(7n^2+20)}{216}\phi^4 \\
&\quad-\frac{n(357n^4+2020n^2+470)}{9720\sqrt{3}}\phi^5
-\frac{n^2(961n^4+8950n^2+6670)}{58320}\phi^6+\cO(\phi^7).
\ea
\label{eq:lambda-cone-pert}
\ee
Note that in this expansion, the correction at order $\phi$ takes the form of $n$ rather than $n+1/2$.
As a consequence, the electron energy behaves as $E_\text{Dirac} \sim 3^{1/4}\sqrt{n \phi}$ near the cone \cite{McClure}.
In figure~\ref{fig:Landau}, we show the Landau level splitting in the $\lambda$-spectrum.
Our perturbative result shows good agreement with the weak magnetic spectrum.

The perturbative expansion \eqref{eq:lambda-cone-pert} is very similar to that in the supersymmetric sine-Gordon
quantum mechanics \cite{DU-zero, KSTU} (see equation \eqref{eq:E-pert-SSG}). In particular, the lowest Landau level eigenvalue does not seem to receive any perturbative corrections:
\be
\lambda_\text{Dirac}^\text{pert}(0,\phi)=-3.
\ee
We have checked this property up to $\phi^{60}$.
As we will see in the next subsections, the nonperturbative structure is also similar. 
Such a structure implies that the honeycomb lattice effectively has hidden supersymmetry near the Dirac point.
This implication is consistent with an earlier work \cite{Ezawa, ESKH}.

\subsection{Nonperturbative bandwidth}
In the previous subsection, we computed the perturbative expansions around two regimes $\lambda=6$ and $\lambda=-3$.
This is not the end of the story.
The spectrum of Bloch electrons in the lattice system under the magnetic field has the subband structure rather than the discrete spectrum.
The naive question is thus how wide these subbands are.
This problem is not simple to answer because in the case of Harper's difference equation,
the spectrum has a self-similar pattern. The subband structure becomes very complicated if the rational flux $\Phi/\Phi_0$ is complicated.

In this paper, we restrict our analysis to the particular case of the flux with the form $\phi=2\pi/Q$, where $Q$ is a large integer.
We consider the weak flux limit $Q \to \infty$, and observe a simple pattern of the width of subbands.
As is expected, the bandwidth turns out to have an exponentially small scale $e^{-\alpha Q}$ in the large $Q$ limit due to quantum mechanical
tunneling effects.
Therefore the analysis of the bandwidth is a first step to understand the nonperturbative effects.
By using the numerical analysis explained in appendix~\ref{sec:num}, we conjecture several analytic forms of the bandwidths.

As an example, we show in table~\ref{tab:band} the numerical values of the edges of subbands of $\lambda$ for $Q=30$.
In this case, there are $30$ subbands, but we show only the five subbands near the top ($\lambda=6$) and near the bottom ($\lambda=-3$), respectively, since we are interested in the bandwidth near these points.
One can observe that these subbands are actually very narrow.
 
\begin{table}[tbp]
\caption{Subband structure of $\lambda$ at $\phi=2\pi/30$. We show a few subbands around the top ($\lambda=6$) and the bottom ($\lambda=-3$).}
\label{tab:band}
\begin{center}
\begin{tabular}{ccc}
\hline
Lower edge & Upper edge & Width \\
\hline
 $5.644554027259942873706$ & $5.644554027259942873729$ & $2.26 \times 10^{-20}$\\
 $4.962033623731763637236$ & $4.962033623731763639303$ & $2.07 \times 10^{-18}$\\
 $4.320798113221331735813$ & $4.320798113221331828144$ & $9.23 \times 10^{-17}$\\
 $3.719185381701187588898$ & $3.719185381701190275939$ & $2.69 \times 10^{-15}$\\
 $3.155584564864439877293$ & $3.155584564864497127576$ & $5.73 \times 10^{-14}$\\
 $\vdots$ & $\vdots$ & $\vdots$ \\
$ -2.368980744700370366683$ & $-2.362300216473043126822$ & $0.00668$\\
 $-2.375181917706344170625$ & $-2.369168804622731871603$ & $0.00601$\\
 $-2.660409627976225867961$ & $-2.659871545251541238427$ & $0.000538$\\
 $-2.660943201255546989412$ & $-2.660410724971488755401$ & $0.000532$\\
 $-3.000000000000000000000$ & $-2.999999998092428049662$ & $1.91 \times 10^{-9}$\\
 \hline
\end{tabular}
\end{center}
\end{table}%

\begin{figure}[t]
\begin{center}
  \begin{minipage}[b]{0.45\linewidth}
    \centering
    \includegraphics[width=0.95\linewidth]{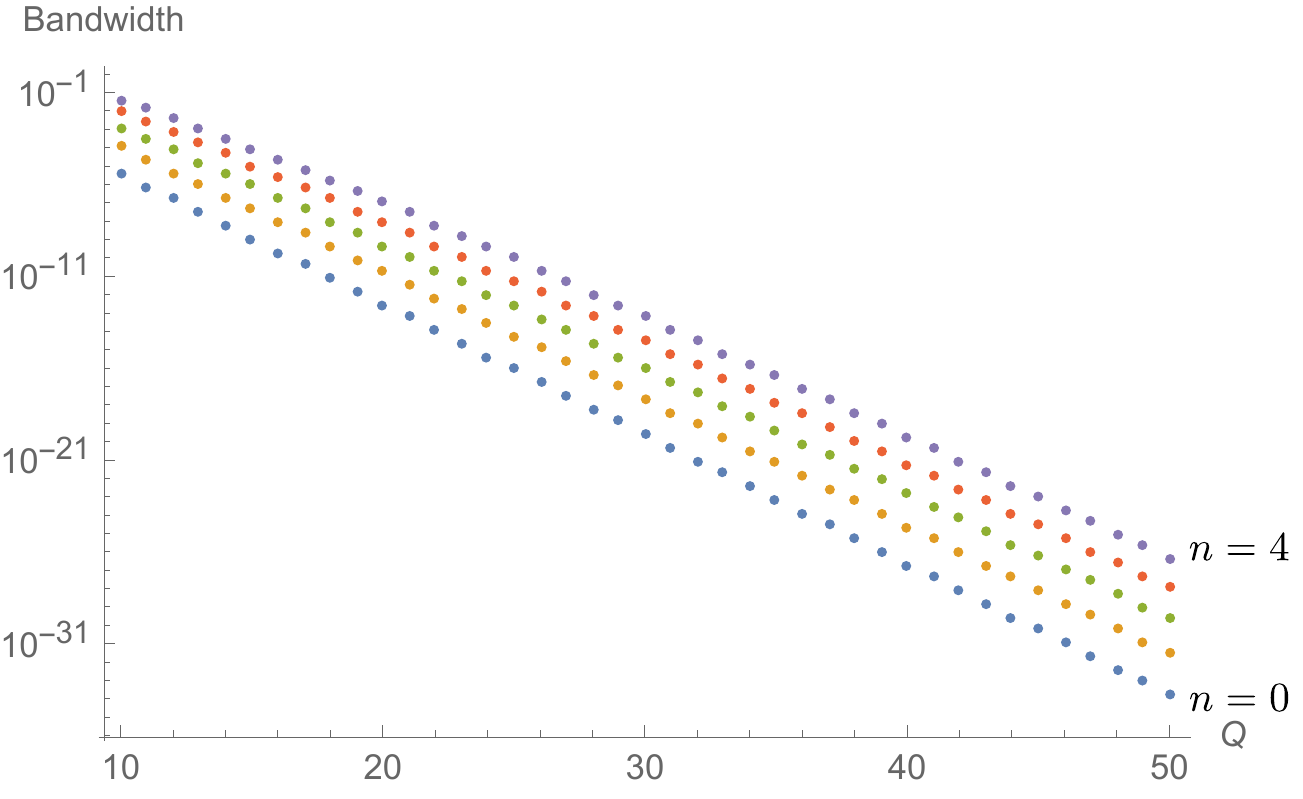}
  \end{minipage} \hspace{1cm}
  \begin{minipage}[b]{0.45\linewidth}
    \centering
    \includegraphics[width=0.95\linewidth]{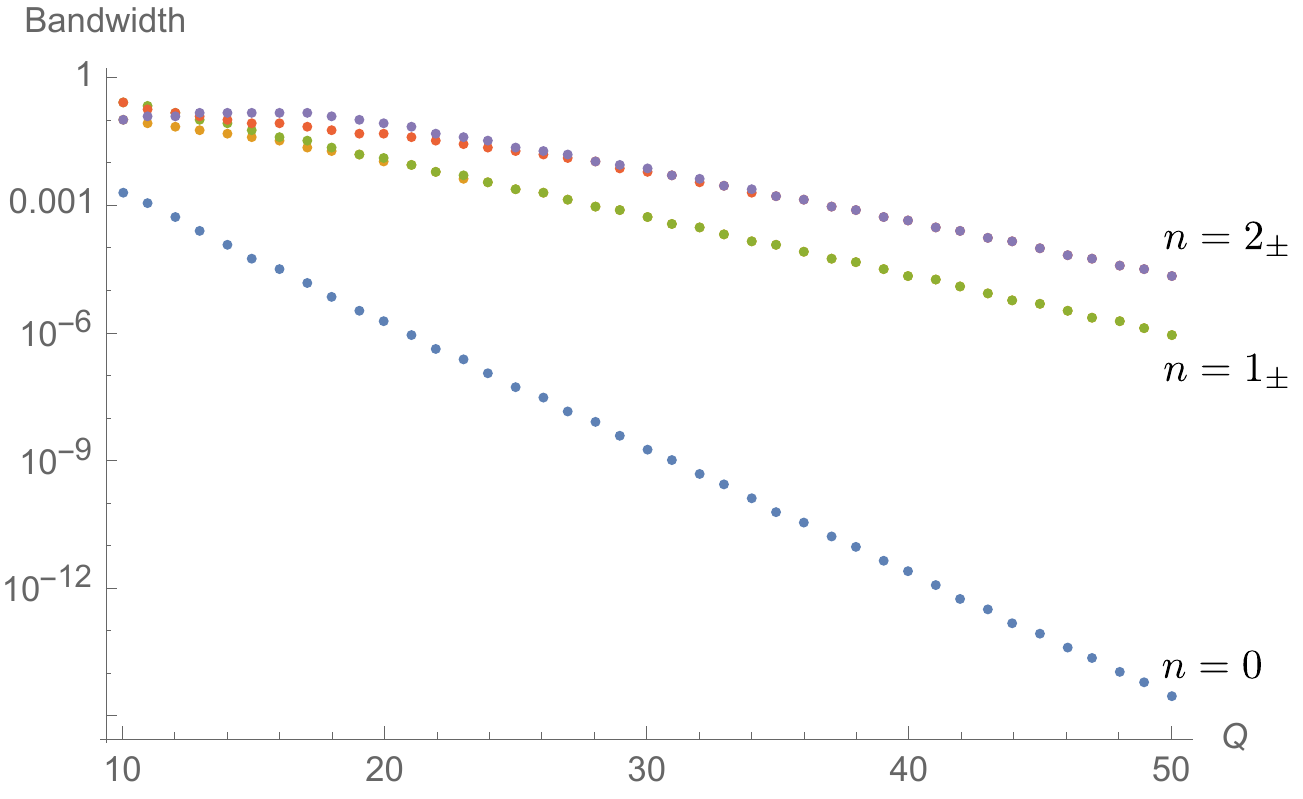}
  \end{minipage} 
\end{center}
  \caption{These figures show the bandwidths near $\lambda=6$ (left) and near $\lambda=-3$ (right) against $Q$.
  In the right figure, there are two subbands whose bandwidths are almost same for each Landau level $n \geq 1$.
  We denote these two by indices $\pm$.}
  \label{fig:width}
\end{figure}

Let us first consider the bandwidth near the top.
In figure~\ref{fig:width} (left), we plot the bandwidths for the first five Landau levels in the range $10 \leq Q \leq 50$.
Clearly, these bandwidths are exponentially small.
We can compute the exponential factor as follows.
As is well-known, the quantum mechanical tunneling effect is explained by the semiclassical WKB period integral for a classically forbidden orbit \cite{LL-textbook}.
In our case, the classical curve of the system is given by \eqref{eq:curve}.
Then the period integral to explain the tunneling effect turns out to be
\be
\exp\biggl[ -\frac{C}{\phi} \int_{-\pi/2}^{\pi/2} dq \, (p^{(+)}_{\lambda=6}(q)-p^{(-)}_{\lambda=6}(q)) \biggr].
\ee
where $p^{(\pm)}_{\lambda}(q)$ are two branches in \eqref{eq:curve}.
By the careful choice of the numerical constant $C$ to match the numerical result, we conclude that the correct exponential factor is
\be
\exp \left[ -\frac{A}{2\pi} Q \right],
\ee
where $A$ is given by
\be
A=\frac{2}{i}\int_{-\pi/2}^{\pi/2} dq \, \arccos\( \frac{2}{\cos q}-\cos q \)=10.149416064\cdots.
\label{eq:A}
\ee 
We also give a conjecture of the prefactor.
By using the numerical fitting of the bandwidths for various Landau levels, we find the leading contribution
\be
\Delta \lambda_\text{top}^\text{band}(n, 2\pi/Q)\approx \frac{108 \cdot 3^{1/4}}{n!} \biggl( \frac{6\sqrt{3}}{\pi} \biggr)^n Q^{n-\frac{1}{2}} e^{-\frac{A}{2\pi}Q}
\cP_\text{top}^\text{inst}(n,2\pi/Q),
\ee
where $\cP_\text{top}^\text{inst}(n,2\pi/Q)=1+\cO(Q^{-1})$ is the perturbative fluctuation around a nonperturbative saddle-point in the path integral perspective. A first few coefficients of $\cP_\text{top}^\text{inst}(n,2\pi/Q)$ will be conjectured in the next subsections.
We expect that it is explained by the one-instanton saddle.
It is straightforward to rewrite this result into the width of the energy bands.

The analysis near the bottom edge is much more complicated. 
We first observe that for each Landau level $n \geq 1$ there is a pair of subbands whose bandwidths are almost same,
as shown in figure~\ref{fig:width} (right).
The gap of these two subbands is extremely narrow, and it is almost regarded as a zero-gap. See table \ref{tab:band} for instance.
As will be seen in the next section, such a zero-gap structure also appears in the supersymmetric sine-Gordon system.
We distinguish these two subbands by subscript $\pm$.
Then we again observe from numerics that each bandwidth of $\lambda$ scales as
\be
\Delta \lambda_\text{Dirac,$\pm$}^\text{band}(n, 2\pi/Q) \approx \frac{3^{\frac{3(n+1)}{2}}\sqrt{n}}{(2\pi)^{n-\frac{1}{2}}n!} Q^{n-1}e^{-\frac{A}{10\pi}Q}
\cP_\text{Dirac, $\pm$}^\text{inst}(n,2\pi/Q),
\qquad n \geq 1,
\ee
where $A$ is the same number given previously.

The lowest Landau level $n=0$ is special. 
As shown in figure~\ref{fig:width} (right), the exponential decay for $n=0$ is much faster than that for $n \geq 1$.
The careful numerical analysis reveals
\be
\Delta \lambda_\text{Dirac}^\text{band}(0, 2\pi/Q) \approx 9\sqrt{3}Q^{-1}e^{-\frac{A}{5\pi} Q} \cP_\text{Dirac}^\text{bion}(0,2\pi/Q) .
\ee
Since the exponential factor is the square of that for $n \geq 1$, the fluctuation $\cP_\text{Dirac}^\text{bion}$ should be explained by
the two-instanton sector. By analogy with quantum mechanics, this contribution comes from the instanton--anti-instanton (i.e. bion) saddle.
Let us translate this result into the energy spectrum.
We use the fact that $\lambda=-3$ is always the bottom edge of the subband at the lowest Landau level.
Hence the top edge of this subband is
\be
\lambda_\text{Dirac}^\text{top edge}(0,2\pi/Q) =-3+\Delta \lambda_\text{Dirac}^\text{band}(0, 2\pi/Q).
\ee
The bandwidth of the lowest Landau energy is finally given by
\be
\Delta E_\text{Dirac}^\text{band}(0,2\pi/Q)=\sqrt{\Delta \lambda_\text{Dirac}^\text{band}(0, 2\pi/Q)}
=3^{5/4} Q^{-1/2}e^{-\frac{A}{10\pi} Q} [ 1+\cO(Q^{-1}) ].
\ee
We conclude that the bandwidth of the energy for the lowest Landau level has the same order as those for $n \geq 1$.

\subsection{Testing a PNP threesome relation}
The analysis in the previous subsection heavily relies on the numerical experiments.
This is a kind of guess works.
We would like to extract universal properties from these data.
Remarkably, recent resurgent analysis shows up new relations between the perturbative sector
and nonperturbative sectors.
Such a resurgent consideration also must tell us about relations among nonperturbative sectors.
For the resurgent analysis in high energy physics, see \cite{Marino:2012zq, Dorigoni:2014hea, Dunne:2015eaa} and references therein. 
Inspired by this development, we here check a simple relation among the perturbative, the one-instanton 
and the instanton--anti-instanton sectors, which is expected by the resurgent analysis in usual quantum mechanics.

We find that the honeycomb lattice has the following beautiful relation
\be
\frac{\cP^\text{bion}(n,\phi)}{\cP^\text{inst}(n,\phi)^2}=\( \frac{1}{\cN}\frac{\pd \lambda^\text{pert}(n,\phi)}{\pd n}\)^{-1} ,
\label{eq:NP-NP-honey}
\ee
where $\cN$ is a normalization factor, which should be chosen as $\cN_\text{Dirac}=\sqrt{3}\phi$ for the Dirac point $\lambda_\text{Dirac}=-3$
and as $\cN_\text{top}=-2\sqrt{3}\phi$ for the top $\lambda_\text{top}=6$. 
For the reader who wonders why this relation is expected, see subsection~\ref{subsec:magic}.
We have neither a rigorous proof nor strong evidence for this relation in the honeycomb system so far.
Nevertheless we believe that \eqref{eq:NP-NP-honey} works both for the Dirac point and for the top.
The reason is that this kind of relations seems to hold very widely not only for well-studied quantum mechanical systems
but also for Hofstadter-type 2d electron systems \cite{ToAppear}.

We here give a few nontrivial checks, based on the numerical analysis.
We focus on the Dirac point.
As was seen  before, the lowest Landau level, the bandwidth starts from the bion contribution.
We observe that its fluctuation is given by
\be
\log \cP_\text{Dirac}^\text{bion}(0,\phi)=-\frac{11\phi}{36\sqrt{3}}-\frac{\phi^2}{27}-\frac{1081\phi^3}{29160\sqrt{3}}+\cO(\phi^4),\qquad
\phi=\frac{2\pi}{Q}.
\label{eq:P-Dirac-bion}
\ee
Also the bandwidth for $n \geq 1$ is explained by the one-instanton correction.
The numerical experiment allows us to find
\be
\ba
\log \cP_\text{Dirac}^\text{inst}(n,\phi)&=-\frac{30n^2+72n+11}{72\sqrt{3}}\phi-\frac{34n^3+96n^2+49n+16}{432}\phi^2 \\
&\quad -\frac{4470n^4+17280n^3+14910n^2+12960n+1081}{58320\sqrt{3}}\phi^3+\cO(\phi^4),
\ea
\label{eq:fluc-Dirac}
\ee
where $\cP_\text{Dirac}^\text{inst}=\cP_\text{Dirac,$+$}^\text{inst}=\cP_\text{Dirac,$-$}^\text{inst}$.
To find this result, we first compute the expansion for various Landau levels $n=1,2,\dots$,
and then determine each coefficient by assuming that the coefficient at order $\phi^k$ is a polynomial of $n$
with degree $k+1$. This assumption is validated by checking whether the obtained result reproduces the ones for higher Landau levels or not. 
Note that the expansion \eqref{eq:fluc-Dirac} is obtained by the analysis for $n \geq 1$, but we can extrapolate it to $n=0$.

Combining these observations, one can evaluate both sides of \eqref{eq:NP-NP-honey} for $n=0$, independently.
The left hand side is
\be
\frac{\cP_\text{Dirac}^\text{bion}(0,\phi)}{\cP_\text{Dirac}^\text{inst}(0,\phi)^2} =1+\frac{\phi^2}{27}+\cO(\phi^4).
\ee
The right hand side is
\be
\biggl(\frac{1}{\sqrt{3}\phi} \frac{\pd \lambda_\text{Dirac}^\text{pert}(n,\phi)}{\pd n}\biggr)^{-1}_{n=0}
=1+\frac{\phi^2}{27}+\cO(\phi^4).
\ee
Indeed these expansions perfectly agree!
To check the relation for $n \geq 1$, we need the bion correction.
This correction can be extracted from the large order behavior of the perturbative expansion.
Let us write the perturbative and the bion expansions as
\be
\lambda_\text{Dirac}^\text{pert}(n,\phi)=\sum_{\ell =0}^\infty a_\ell^{(0)}(n) \phi^\ell,\qquad
\cP_\text{Dirac}^\text{bion}(n,\phi)=\sum_{\ell=0}^\infty a_\ell^{(1,1)}(n) \phi^\ell,
\ee
where our normalization is $a_{0}^{(0)}(n)=-3$ and $a_0^{(1,1)}(n)=1$.
The resurgent analysis (see \cite{Marino:2012zq, Dorigoni:2014hea, Dunne:2015eaa} for example) tells us that the information on $a_\ell^{(1,1)}(n)$ is encoded in the large order behavior of $a_\ell^{(0)}(n)$.
The large order behavior has the following form
\be
\ba
a_\ell^{(0)}(n)=-\frac{C_n}{2\pi} \frac{(\ell+2n-2)!}{(2\cS)^{\ell+2n-1}} \biggl[ 1+\frac{2\cS a_1^{(1,1)}(n)}{\ell+2n-2}
+\frac{(2\cS)^2 a_2^{(1,1)}(n)}{(\ell+2n-2)(\ell+2n-3)} \\
+\frac{(2\cS)^3 a_3^{(1,1)}(n)}{(\ell+2n-2)(\ell+2n-3)(\ell+2n-4)}+\cdots \biggr]+\cdots
\ea
\ee
in the limit $\ell \to \infty$. Here $\cS=A/5$ and $C_n$ is an $n$-dependent constant.
We find $C_1=-81\sqrt{3}$ and $C_2=-2187\sqrt{3}/2$.
We have the perturbative data $a_\ell^{(0)}(n)$ ($n=1,2$) up to $\ell=100$, and use them to estimate $a_\ell^{(1,1)}(n)$ ($n=1,2$) numerically by the method in appendix~\ref{sec:num}.
We then obtain the following numerical values of the bion corrections:
\be
\ba
a_1^{(1,1)}(1) &\approx -1.2348880757, \quad a_2^{(1,1)}(1) \approx 0.1189557619, \quad 
a_3^{(1,1)}(1) \approx -0.2236656317, \\
a_1^{(1,1)}(2) &\approx -3.2556140204, \quad a_2^{(1,1)}(2) \approx 2.6606223304, \quad 
a_3^{(1,1)}(2) \approx -1.4032762609,
\ea
\label{eq:a11-num}
\ee
Now, we compare these values with the prediction from the relation \eqref{eq:NP-NP-honey}.
Using the expansions of $\lambda_\text{Dirac}^\text{pert}$ and $\cP_\text{Dirac}^\text{inst}$, 
one easily obtains the fluctuation around the bion:
\be
\ba
\cP_\text{Dirac}^\text{bion}(1,\phi)&=1-\frac{77}{36\sqrt{3}}\phi+\frac{925}{7776}\phi^2-\frac{1626709}{4199040\sqrt{3}}\phi^3+\cO(\phi^4), \\
\cP_\text{Dirac}^\text{bion}(2,\phi)&=1-\frac{203}{36 \sqrt{3}} \phi +\frac{20689}{7776} \phi ^2-\frac{10205959}{4199040 \sqrt{3}} \phi ^3+\cO(\phi^4).
\ea
\ee
The coefficients are actually in agreement with the numerical estimation \eqref{eq:a11-num}.

For the nonperturbative corrections around the top, it is not easy to check the relation because the bion fluctuation
is hard to be evaluated.%
\footnote{As was done above, such a bion fluctuation is captured by the large order behavior of the perturbative expansion,
but in this case it is problematic to extract it. We postpone this issue to future works.}
Alternatively, we use \eqref{eq:NP-NP-honey} to predict the bion correction.
For the top, we find the following one-instanton fluctuation
\be
\ba
\log \cP_\text{top}^\text{inst}(n,\phi)&=-\frac{6n^2+42n+19}{72\sqrt{3}}\phi-\frac{2n^3+15n^2+15n+6}{864}\phi^2 \\
&\quad-\frac{15n^4+138n^3+258n^2+297n+166}{46656\sqrt{3}}\phi^3+\cO(\phi^4) .
\ea
\ee
Our threesome relationship predicts the bion fluctuation:
\be
\ba
\log \cP_\text{top}^\text{bion}(n,\phi)&=-\frac{3n^2+12n+5}{18\sqrt{3}}\phi
-\frac{4n^3+18n^2+18n+7}{864}\phi^2 \\
&\quad -\frac{30n^4+168n^3+354n^2+378n+251}{46656\sqrt{3}}\phi^3+\cO(\phi^4).
\ea
\ee
It would be nice to check whether this prediction is correct or not.

\section{Remarks on SUSY sine-Gordon QM}\label{sec:SSG}
In this section, we give some remarks on nonperturbative effects of the supersymmetric sine-Gordon quantum mechanics.
In \cite{DU-zero, KSTU, Keio}, the instanton--anti-instanton correction was mainly studied in the context of resurgence theory.
Here we look into the bandwidth of this model, which is not discussed in these references.
The bandwidth is caused by one-instanton effect.

\subsection{Band structure}
We start with the following supersymmetric quantum mechanics:
\be
H_\text{SQM}=-\frac{g}{2}\frac{d^2}{dx^2}+\frac{1}{2g}(W'(x)^2+g W''(x) \sigma_3),
\ee
where $\sigma_i$ ($i=1,2,3$) are the Pauli matrices, and $W(x)$ is a superpotential.
The first component corresponds to the fermionic sector, while the second component to the bosonic sector.
We here consider the sine-Gordon potential:
\be
W(x)=-\cos x.
\ee  
For our purpose, it is convenient to rescale the variable by $x=\sqrt{g}q$.
Then, the bosonic sector of the Hamiltonian is
\be
H=-\frac{1}{2}\frac{d^2}{dq^2}+\frac{1}{2g} \sin^2(\sqrt{g}q)-\frac{1}{2}\cos (\sqrt{g}q).
\ee
This Hamiltonian admits the expansion around $g=0$,
\be
H=-\frac{1}{2}\frac{d^2}{dq^2}+\frac{q^2}{2}-\frac{1}{2}+g\(\frac{q^2}{4}-\frac{q^4}{6} \)+g^2\(-\frac{q^4}{48}+\frac{q^6}{45}\)+\cO(g^3).
\ee
We can use the Bender-Wu method to compute the perturbative expansion of the energy.
Using the \textit{Mathematica} package in \cite{SU-BW}, we obtain
\be
\ba
E^\text{pert}(n,g)&=n-\frac{n^2}{4}g-\frac{n(2n^2+1)}{32}g^2-\frac{n^2(5n^2+7)}{128}g^3 \\
&\quad-\frac{n(66n^4+182n^2+25)}{2048}g^4-\frac{n^2(63n^4+290n^2+127)}{2048}g^5+\cO(g^6).
\ea
\label{eq:E-pert-SSG}
\ee
As was mentioned before, this perturbative expansion is very similar to \eqref{eq:lambda-cone-pert}.
In particular, the expansion exactly vanishes for $n=0$:
\be
E^\text{pert}(0,g)=0.
\ee

\begin{figure}[t]
\begin{center}
  \begin{minipage}[b]{0.45\linewidth}
    \centering
    \includegraphics[width=0.95\linewidth]{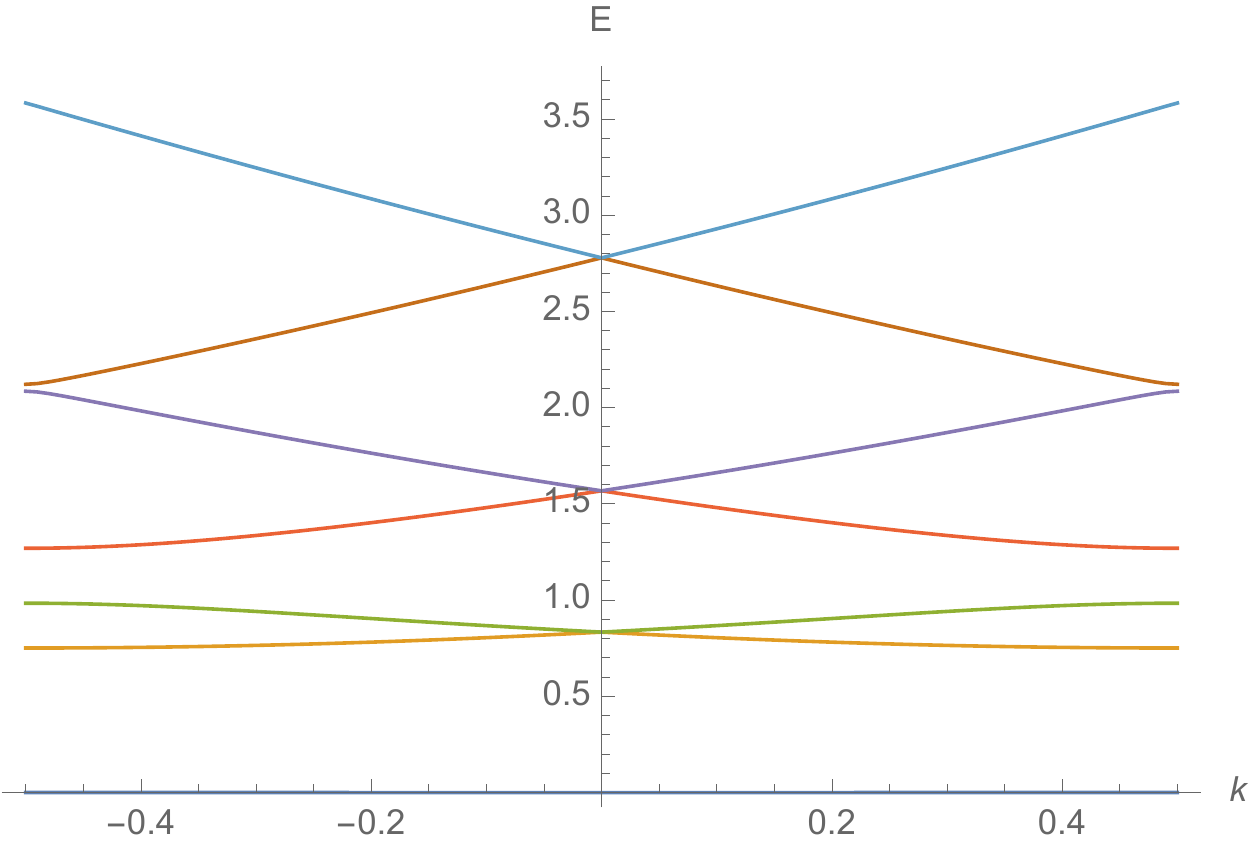}
  \end{minipage} \hspace{1cm}
  \begin{minipage}[b]{0.45\linewidth}
    \centering
    \includegraphics[width=0.95\linewidth]{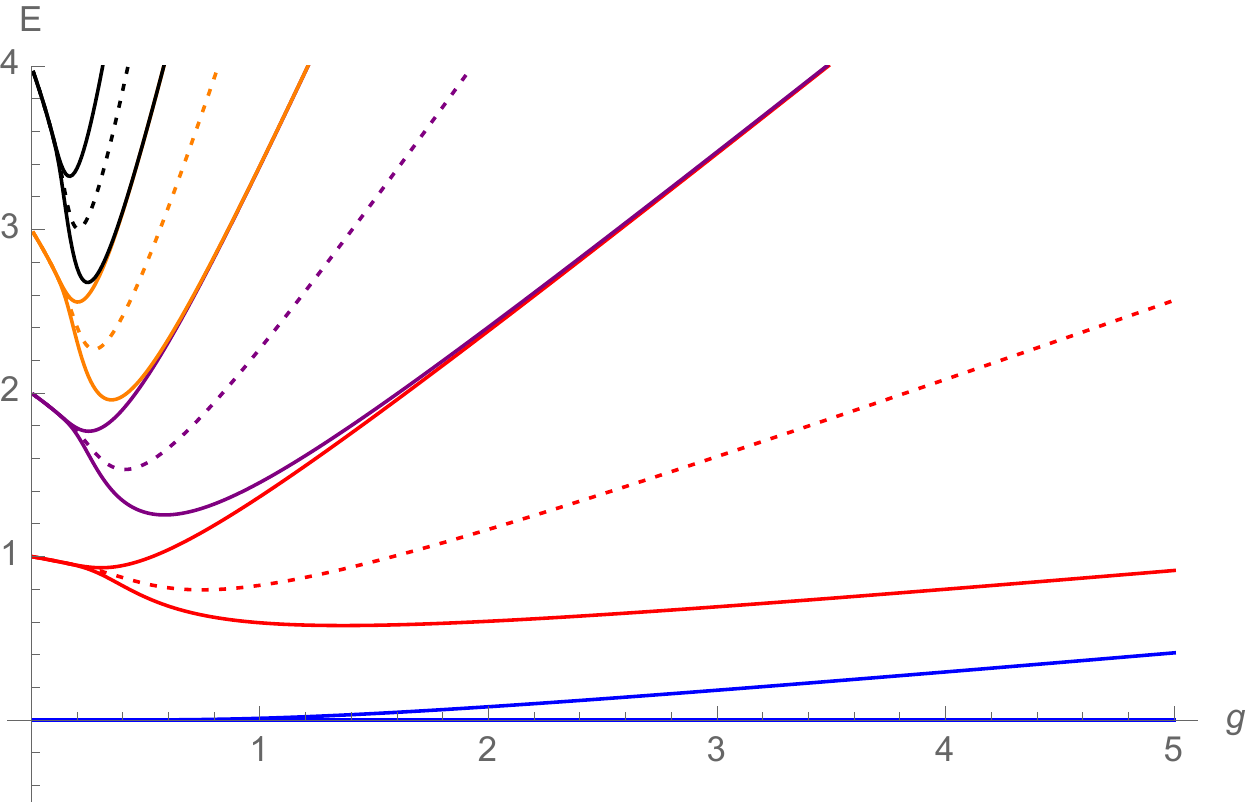}
  \end{minipage} 
\end{center}
  \caption{The left figure shows the first seven energies at $g=1/2$ against the wave number in the first Brillouin zone. There is a zero-gap at $k=0$ between the $(2m-1)$-th band and the $2m$-th band. The right figure shows the coupling dependence of the energies. The solid lines represent the edges of bands. The dotted lines represent the zero-gap energies. Note that $E=0$ is always the lowest energy due to supersymmetry.}
  \label{fig:SSG}
\end{figure}

Let us see the band structure.
Since the supersymmetric sine-Gordon system has a $2\pi$-periodic potential, 
the Floquet-Bloch theorem states that the wave function takes the form of
\be
\psi_k(x)=e^{ikx} \sum_{n \in \mathbb{Z}} c_n e^{inx},
\ee 
where $k$ is a continuous parameter, ranging from $k=-1/2$ to $k=1/2$.
Plugging this into the Schr\"odinger equation, we obtain an infinite set of relations
\be
\left[ \frac{g}{2}(n+k)^2+\frac{1}{4g} \right]c_n-\frac{1}{4}(c_{n-1}+c_{n+1})-\frac{1}{8g}(c_{n-2}+c_{n+2})=E_k(g) c_n
\label{eq:eigen-eq}
\ee
By diagonalizing the left hand side of this equation, we obtain the discrete eigenvalues $E_k(m, g)$ ($m=0,1,2,\dots$) 
for a given wave number $k$ and coupling $g$.
Since the wave number moves in the first Brillouin zone ($-1/2 \leq k \leq 1/2$) continuously, the spectrum forms energy bands.
We show the $k$-dependence of $E_k(g)$ for $g=1/2$ in figure~\ref{fig:SSG} (left).

One characteristic feature of this model is that the upper band edge with band index $2m-1$ 
and the lower band edge with band index $2m$ coincide for all $m \geq 1$. This means that there are an infinite number of zero-gap energies:
$E_{k=0}(2m-1,g)=E_{k=0}(2m,g)$.
For convenience we denote these bands as
\be
E_{k,-}(m,g):=E_k(2m-1, g), \qquad E_{k,+}(m,g):=E_k(2m, g),\qquad m\geq 1.
\ee
Another important fact is that $E=0$ is always the lowest eigenvalue at $k=0$: $E_{k=0}(0,g)=0$.
This is a reflection that the system has supersymmetry.
The ground state does not receive any quantum corrections.
In figure~\ref{fig:SSG} (right), we plot the energy as a function of the coupling $g$.

Let us proceed to the computation of the bandwidth.
We do it numerically.
The bandwidth is given by
\be
\Delta E^\text{band}_\pm (n,g):=|E_{k=0, \pm}(n,g)-E_{k=1/2,\pm}(n,g)|
\ee
For the lowest band $n=0$, we also define $\Delta E^\text{band}(0,g):=E_{k=1/2}(0,g)$.
We first solve the eigenvalue equation \eqref{eq:eigen-eq} for various $g$ numerically.
We then fit these values under the ansatz
\be
\log \Delta E^\text{band}_\pm (n,g) \approx -\frac{A}{g}+B\log g+C+\sum_{j=1}^\infty c_j g^j.
\ee
The numerical fitting allows us to guess exact values of $A$, $B$ and $C$.
After that, we use the method in appendix~\ref{sec:num} to determine $c_j$'s.

For the lowest band, we find
\be
\Delta E^\text{band}(0,g)\approx \frac{2}{\pi}e^{-\frac{4}{g}}\cP^\text{bion}(0,g),
\ee
where $\cP^\text{bion}(0,g)$ is expected to come from the bion saddle in the path integral formulation.
We observe that it has the following perturbative expansion 
\be
\cP^\text{bion}(0,g)=1-\frac{g}{8}-\frac{3g^2}{128}-\frac{13g^3}{1024}-\frac{341g^4}{32768}+\cO(g^5).
\label{eq:bion-fluc}
\ee
For $n \geq 1$, we also find the leading correction
\be
\Delta E^\text{band}_\pm (n,g) \approx \frac{2^{n+\frac{1}{2}}}{n!} \sqrt{\frac{n}{\pi}} \( \frac{4}{g} \)^n e^{-\frac{2}{g}} \cP^\text{inst}(n,g),
\ee
where the fluctuation $\cP^\text{inst}(n,g)$ does not depend on two $\pm$-bands, and it should be explained by the one-instanton saddle (and the one-anti-instanton saddle).
We can guess a few terms
\be
\ba
\log \cP^\text{inst}(n,g) &=-\frac{6n^2+8n+1}{16}g-\frac{10n^3+20n^2+7n+2}{64}g^2 \\
&\quad-\frac{330n^4+896n^3+546n^2+384n+25}{3072}g^3+\cO(g^4).
\ea
\label{eq:1inst-fluc}
\ee
Note that the bandwidth for $n=0$ comes from the bion contribution $\cO(e^{-4/g})$, while for $n \geq 1$ from the one-instanton contribution $\cO(e^{-2/g})$.
This property is the same as the bandwidth of $\lambda$ for the honeycomb lattice in section~\ref{sec:pert}.
The similar behavior is also found in the supersymmetric double well potential (or the Fokker-Planck model in the literature) \cite{ZJJ3}. 

\subsection{PNP relations}\label{subsec:magic}
There is a remarkable relationship between the perturbative sector and the nonperturbative sectors.
We refer to it as the perturbative/nonperturbative (PNP) relation \cite{AC1, AC2, AHS, Alvarez} (see also \cite{DU0, DU1, DU-Mathieu}).
In our case, this relation results in the form
\be
\cP^\text{inst}(n,g)=\frac{\pd E^\text{pert}(n,g)}{\pd n}
\exp \biggl[ S^\text{inst} \int_0^g \frac{dg}{g^2}\biggl( \frac{\pd E^\text{pert}(n,g)}{\pd n}-1+\frac{ng}{S^\text{inst}} \biggr) \biggr],
\label{eq:DU-1}
\ee
where $S^\text{inst}=2$ is the instanton action.
We note that the similar formula was found long time ago \cite{Stark}.
Since the right hand side in this equation contains only the perturbative quantity,
the one-instanton correction is completely controlled by the perturbative sector!
Using the perturbative expansion \eqref{eq:E-pert-SSG}, one can check that it reproduces our guess \eqref{eq:1inst-fluc} based on the numerical analysis.

The authors in \cite{DU-zero, KSTU} conjectured that the fluctuation around the bion saddle also has a similar formula:
\be
\cP^\text{bion}(n,g)=\frac{\pd E^\text{pert}(n,g)}{\pd n}
\exp \biggl[ 2S^\text{inst}\int_0^g \frac{dg}{g^2}\biggl( \frac{\pd E^\text{pert}(n,g)}{\pd n}-1+\frac{ng}{S^\text{inst}} \biggr) \biggr].
\label{eq:DU-2}
\ee
This conjecture indeed reproduces \eqref{eq:bion-fluc} for $n=0$.
Also, the conjecture was confirmed by comparing it with the large order prediction \cite{KSTU}.

Obviously, these two equations are almost the same form.
After a simple computation, we arrive at
\be
\frac{\cP^\text{bion}(n,g)}{\cP^\text{inst}(n,g)^2}=\(\frac{\pd E^\text{pert}(n,g)}{\pd n}\)^{-1}.
\label{eq:NP-NP-2}
\ee
This is just the formula we saw in the previous section.

In well-studied examples such as the double well potential and the cosine potential, this relation is not so surprising,
because in these cases, we already know that the Zinn-Justin--Jentschura (ZJJ) exact quantization conditions \cite{ZJJ1, ZJJ2}
contain only the two non-trivial functions,
a ``perturbative'' function $B(E, g)$ and a ``nonperturbative'' function $A(E,g)$.
The complete trans-series expansion of the spectrum is expressed in terms of this two functions in principle.
Moreover, these two functions $A(E,g)$ and $B(E, g)$ are exactly related by the PNP relation \cite{AC1, AC2, AHS, Alvarez}.%
\footnote{Note that for quantum mechanical systems with genus more than one classical curves, the naive application of this resurgent relation does not work \cite{Gahramanov:2015yxk}. However, it is expected that there are still generalized resurgent relations even for these cases \cite{BDU}.}
On one hand, the formulae \eqref{eq:DU-1} and \eqref{eq:DU-2} are understood as a consequence of the
PNP relation.
On the other hand, the relation \eqref{eq:NP-NP-2} is considered to be a direct consequence of the ZJJ exact quantization conditions.

A remarkable point is that the relation \eqref{eq:NP-NP-2} seems to hold universally.
As will be reported in \cite{ToAppear}, it also works in the Hofstadter model \cite{Hof}.
In particular, we are not able to find the one-instanton formula like \eqref{eq:DU-1}
for the Hofstadter model nor for the honeycomb lattice model so far. 
It seems to be problematic because in these cases the instanton action $S^\text{inst}=A$ is a complicated irrational number but the coefficients
of the fluctuation are simple rational numbers.
Nevertheless we have observed in the previous section that the threesome relation \eqref{eq:NP-NP-honey} is very likely true even for the honeycomb lattice.

\section{Concluding remarks}\label{sec:conclusion}
In this paper, we analyzed the Bloch electrons in the honeycomb lattice under the magnetic flux.
This system is physically important since graphene has the honeycomb structure. The system has a singular behavior at the Dirac points.
We presented a systematic way to push the computation of the weak magnetic flux expansion perturbatively.
We also looked into the nonperturbative bandwidth of the spectrum, and found a similarity to the supersymmetric sine-Gordon system.
This fact suggests that physics near the Dirac points in graphene has a hidden supersymmetric structure (see also \cite{Ezawa}).%
\footnote{Note that the Dirac cone also exists in the Hofstadter model at $\phi=\pi$ \cite{RP}.}
The nonperturbative effects there is probably related to ``Cheshire Cat Resurgence'' \cite{KSTU, Dorigoni:2017smz}.

Based on the detailed numerical analysis of the nonperturbative corrections, we finally found a simple threesome relation
among the perturbative, one-instanton and instanton--anti-instanton sectors.
For simple quantum mechanical systems, this should be understood as a consequence of the ZJJ exact quantization conditions \cite{ZJJ1, ZJJ2}.
The remarkable point is that this relation seems to be true even for the 2d Bloch electron systems, in which
neither the ZJJ quantization conditions nor the PNP relation have been understood.
These observations suggest that the instanton fluctuation and the bion fluctuation are universally written as
\be
\ba
\cP_\text{fluc}^\text{inst}(n,g)&=\frac{\pd \cP_\text{fluc}^\text{vac}(n,g)}{\pd n} e^{-\widetilde{A}(n,g)}, \\
\cP_\text{fluc}^\text{bion}(n,g)&=\frac{\pd \cP_\text{fluc}^\text{vac}(n,g)}{\pd n} e^{-2\widetilde{A}(n,g)},
\ea
\ee
where $\widetilde{A}(n,g)$ is a function that is essentially same as the function $A(E(n),g)$ appearing in the ZJJ quantization conditions.
It is not necessary to impose the PNP relation between $\cP_\text{fluc}^\text{vac}(n,g)$ and $A(n,g)$ to derive \eqref{eq:NP-NP}.
Our relation might give a clue to show up the nonperturbative structure in the Bloch electron systems.
We gave a piece of evidence of the relation \eqref{eq:NP-NP} for the honeycomb lattice.

We propose several issues. First, it is desirable to understand \eqref{eq:NP-NP} more deeply in Bloch electron systems.
We will investigate it for the Hofstadter model in more detail \cite{ToAppear}.
Second, it would be interesting to ask what quantum geometry of toric Calabi-Yau corresponds to this honeycomb system.
We will report this issue in \cite{ToAppear1}.
In this direction, it is also interesting to unifying the resurgent analysis with the topological string analysis, 
along the line in \cite{CSESV-1, CSESV-2, CSMS, CM, CMS}.
Finally, in our analysis, the similarity of the honeycomb lattice and the SUSY sine-Gordon QM was found.
As in \cite{ESKH}, if introducing an anisotropic parameter in the honeycomb lattice, it leads to a SUSY breaking. 
Does it has a relation to the deformation in \cite{KSTU, Keio}? 
An interesting point is that for particular choices of the parameter in \cite{KSTU, Keio}, the system 
reduces to a quasi-exact solvable system. It would be interesting to look for such deformed electron systems.

\acknowledgments{
We thank Zhihao Duan, Gerald Dunne, Toshiaki Fujimori, Jie Gu, Masazumi Honda, Etsuko Itou, Hosho Katsura, Yuji Sugimoto, Tin Sulejmanpasic, Yuya Tanizaki, Mithat \"Unsal and Zhaojie Xu for discussions. We are especially grateful to Gerald Dunne for many helpful comments on the manuscript. We are also grateful to Marcos Mari\~no for pointing out many relevant references. This work is supported by Rikkyo University Special Fund for Research.
}

\appendix

\section{Numerical method}\label{sec:num}
Our approach in this paper is to decode nonperturbative corrections from numerical data.
In this appendix, we explain how to do so with high precision.
Our method here is based on an idea in \cite{MSW}.
The basic tool is the (generalized) Richardson transform.
The Richardson transform we use here is
\be
\cR_m [f_n]:=\sum_{k=0}^m \frac{ (-1)^{k+m}(n+k)^m}{k! (m-k)!}f_{n+k}.
\ee
where $f_n$ is a given sequence.
This Richardson transform accelerates the speed of convergence for logarithmically convergent sequences.
In fact, if $f_n$ converges as
\be
f_n=C [1+\cO(n^{-1}) ], \qquad n \to \infty,
\ee
then its Richardson transform is
\be
\cR_m [f_n]=C [1+\cO(n^{-m-1}) ], \qquad n \to \infty.
\ee
Therefore the convergence speed is actually improved.
In practical computations, we carefully choose $m$ as good convergence as possible because we have a finite number of $f_n$.

\begin{figure}[t]
\begin{center}
  \begin{minipage}[b]{0.45\linewidth}
    \centering
    \includegraphics[width=0.95\linewidth]{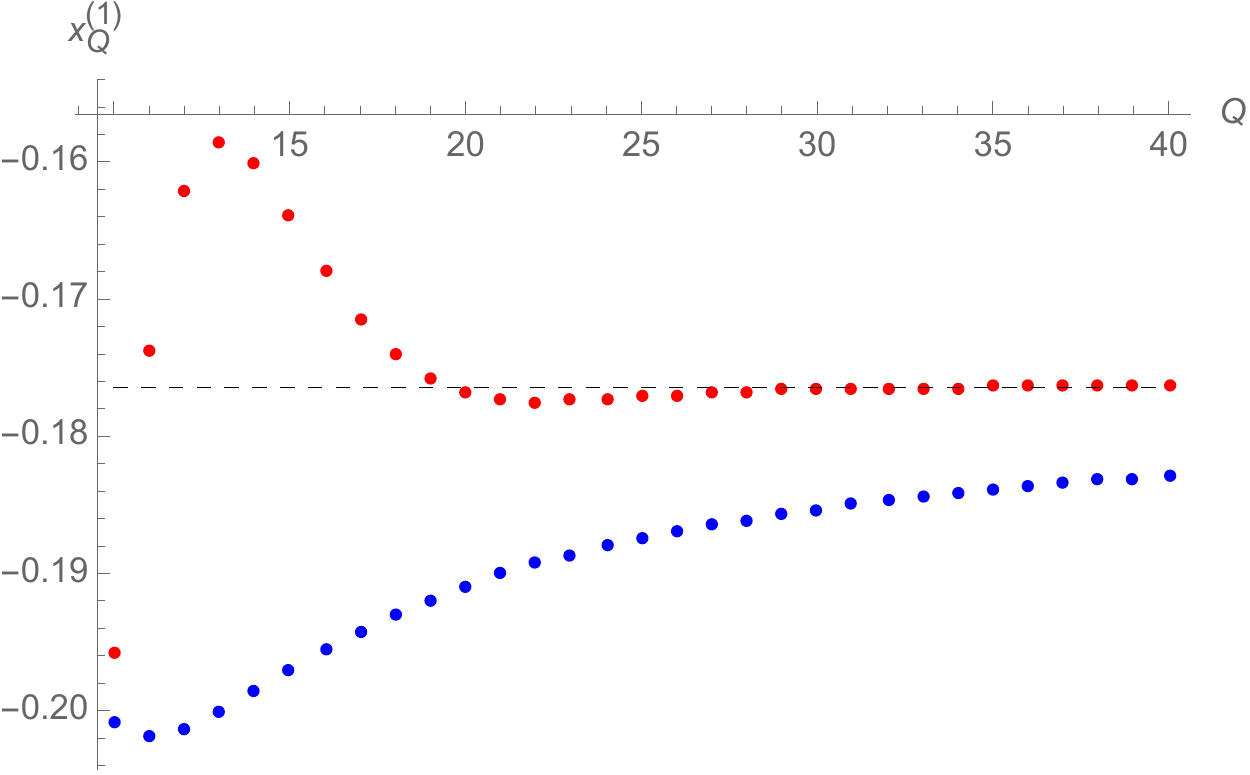}
  \end{minipage} \hspace{1cm}
  \begin{minipage}[b]{0.45\linewidth}
    \centering
    \includegraphics[width=0.95\linewidth]{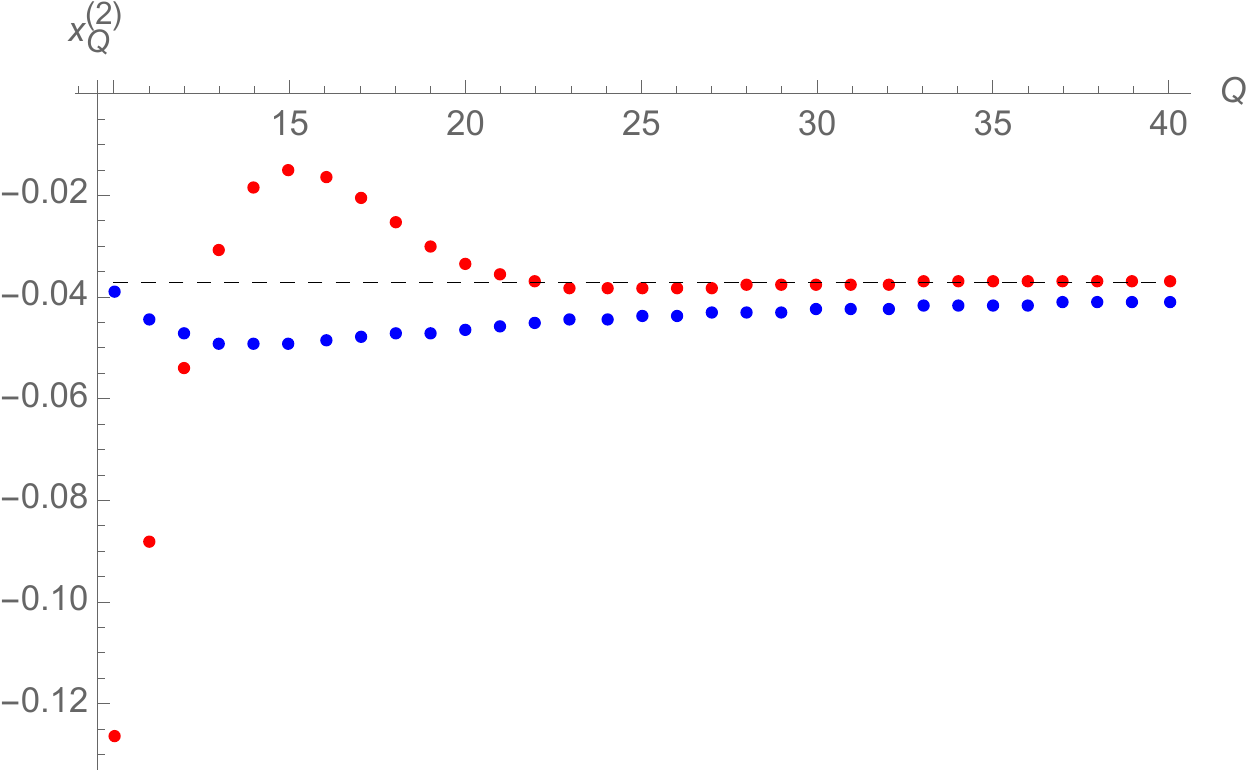}
  \end{minipage} 
\end{center}
  \caption{The Richardson transform accelerate the convergence of sequence. We show the results for $x_Q^{(1)}$ (left) and $x_Q^{(2)}$ (right). The blue dots represent the original sequence and the red dots represent its fifth Richardson transform. The dashed line is the expected convergent value in $Q \to \infty$.}
  \label{fig:Richardson}
\end{figure}

Let us see concrete examples. We show how to determine the bion fluctuation \eqref{eq:P-Dirac-bion}.
Let us define a sequence
\be
x_Q^{(1)}:=\frac{Q}{2\pi} \log \biggl[ \frac{Q}{9\sqrt{3}} e^{\frac{A}{5\pi}Q}\Delta \lambda_\text{Dirac}^\text{band}(0,2\pi/Q) \biggr].
\ee
We assume the behavior $x_Q^{(1)}=x_{\infty}^{(1)}+\cO(1/Q)$ in $Q \to \infty$, where $x_\infty^{(1)}$ is a constant that we want to know.
The convergence speed of this sequence is actually very slow, and it is not possible to guess its convergent value $x_\infty^{(1)}$
from $x_Q^{(1)}$ directly.
The Richardson transform resolves this problem.
In figure~\ref{fig:Richardson} (left), we show the behavior of $x_Q^{(1)}$ and its fifth Richardson transform $\cR_5[x_Q^{(1)}]$ for $10 \leq Q \leq 40$.
It is obvious to see that $\cR_5[x_Q^{(1)}]$ converges much more rapid than the original sequence $x_Q^{(1)}$.

We computed the bandwidth $\Delta \lambda_\text{Dirac}^\text{band}(0,2\pi/Q)$ up to $Q=250$, and applied the 35th Richardson transform.
Then we obtained the following numerical value
\be
\sqrt{3} \cR_{35}[x_{215}^{(1)}]=-0.305555555555555555555555555555558599\cdots,
\ee
whose analytic value is easily guessed as $-11/36$. In this way, we get the analytic value
\be
x_\infty^{(1)}=-\frac{11}{36\sqrt{3}}.
\ee
The next-to-leading coefficient is also obtained by the sequence
\be
x_Q^{(2)}:=\(\frac{Q}{2\pi}\)^2 \biggl[  \log \biggl( \frac{Q}{9\sqrt{3}} e^{\frac{A}{5\pi}Q}\Delta \lambda_\text{Dirac}^\text{band}(0,2\pi/Q) \biggr)
+\frac{11}{36\sqrt{3}} \frac{2\pi}{Q} \biggr].
\ee
The behaviors of $x_Q^{(2)}$ and its fifth Richardson transform are shown in figure~\ref{fig:Richardson} (right).
By the similar way above, we obtain the numerical value 
\be
\cR_{35}[x_{215}^{(2)}]=-0.0370370370370370370370370370373283\cdots.
\ee
It is easy to find $x_\infty^{(2)}=-1/27$.

In the last step to guess analytic values of the coefficients, the function {\tt Rationalize} in \textit{Mathematica} is useful. 
Alternatively, there is a very helpful website \cite{EZ-face} to find complicated irrational numbers.
Repeating this method, we found many analytic values, as in the main text.


\begin{thebibliography}{99}


\bibitem{Hof}  
D.~R.~Hofstadter,
``Energy levels and wave functions of Bloch electrons in rational and irrational magnetic fields,''
Phys. Rev. B {\bf 14} (1976) 2239.  

\bibitem{HKT} 
  Y.~Hatsuda, H.~Katsura and Y.~Tachikawa,
  ``Hofstadter's butterfly in quantum geometry,''
  New J.\ Phys.\  {\bf 18}, no. 10, 103023 (2016)
  [arXiv:1606.01894 [hep-th]].
  
\bibitem{GHM1} 
  A.~Grassi, Y.~Hatsuda and M.~Marino,
  ``Topological Strings from Quantum Mechanics,''
  Annales Henri Poincare {\bf 17}, no. 11, 3177 (2016)
  [arXiv:1410.3382 [hep-th]].
  
  

\bibitem{HSX} 
  Y.~Hatsuda, Y.~Sugimoto and Z.~Xu,
  ``Calabi-Yau geometry and electrons on 2d lattices,''
  Phys.\ Rev.\ D {\bf 95}, no. 8, 086004 (2017)
  [arXiv:1701.01561 [hep-th]].
  
\bibitem{CW}
F.~H.~Claro and G.~H.~Wannier,
``Magnetic subband structure of electrons in hexagonal lattices,''
Phys. Rev. B {\bf 19} (1979) 6068.  

\bibitem{Rammal}
R.~Rammal, ``Landau level spectrum of Bloch electrons in a honeycomb lattice,''
J. Phys. France {\bf 46} (1985) 1345--1354.
  
\bibitem{ToAppear1}
Y.~Hatsuda, Y.~Sugimoto and Z.~Xu, in progress.



  
 \bibitem{McClure}
 J.~W.~McClure, ``Diamagnetism of Graphite,''
 Phys. Rev. {\bf 104} (1956) 666--671.
 
  
\bibitem{BW}
C.~M.~Bender and T.~T.~Wu,
``Anharmonic Oscillator,''
Phys. Rev. {\bf 184} (1969) 1231--1260.

\bibitem{RB}
R.~Rammal and J.~Bellissard,
``An algebraic semi-classical approach to Bloch electrons in a magnetic field,''
J. Phys. France {\bf 51} (1990) 1803--1830. 

\bibitem{ESKH}
K.~Esaki, M.~Sato, M.~Kohmoto and B.~I.~Halperin,
``Zero modes, energy gap, and edge states of anisotropic honeycomb lattice in a magnetic field,''
Phys. Rev. B {\bf 80} (2009) 125405, [arXiv:0906.5027 [cond-mat.mes-hall]].

\bibitem{DU-zero} 
  G.~V.~Dunne and M.~\"Unsal,
  ``Deconstructing zero: resurgence, supersymmetry and complex saddles,''
  JHEP {\bf 1612}, 002 (2016)
  [arXiv:1609.05770 [hep-th]].

\bibitem{KSTU} 
  C.~Koz\c caz, T.~Sulejmanpasic, Y.~Tanizaki and M.~\"Unsal,
  ``Cheshire Cat resurgence, Self-resurgence and Quasi-Exact Solvable Systems,''
  arXiv:1609.06198 [hep-th].
    
  
  \bibitem{AC1}
G.~\'{A}lvarez and C.~Casares, ``Uniform asymptotic and JWKB expansions for anharmonic oscillators,'' J. Phys. A{\bf 33} (2000) 2499.  
  
  
\bibitem{AC2}  
G.~\'{A}lvarez and C.~Casares, ``Exponentially small corrections in the asymptotic expansion of the eigenvalues of the cubic anharmonic oscillator,'' J.~Phys.~A{\bf 33} (2000) 5171.  
  
\bibitem{AHS}  
G.~\'{A}lvarez, C.~J.~Howls and H.~J.~Silverstone, ``Anharmonic oscillator discontinuity formulae up to second-exponentially-small order,'' J. Phys. A{\bf 35} (2002) 4003.  
  
  
\bibitem{Alvarez}
G.~\'Alvarez, ``Langer-Cherry derivation of the multi-instanton expansion for the symmetric double well'', 
J. Math. Phys. {\bf 45}, 3095 (2004).




 

\bibitem{ToAppear}
Z.~Duan, J.~Gu, Y.~Hatsuda and T.~Sulejmanpasic, in progress.



\bibitem{SU-BW} 
  T.~Sulejmanpasic and M.~\"Unsal,
  ``Aspects of Perturbation theory in Quantum Mechanics: The BenderWu Mathematica package,''
  arXiv:1608.08256 [hep-th].

\bibitem{GS-BW}
  J.~Gu and T.~Sulejmanpasic,
  ``High order perturbation theory for difference equations and Borel summability of quantum mirror curves,''
  arXiv:1709.00854 [hep-th].
  
  \bibitem{Ezawa} 
  M.~Ezawa,
  ``Supersymmetry and unconventional quantum Hall effect in graphene,''
  Phys.\ Lett.\ A {\bf 372}, 924 (2008)
  [cond-mat/0606084 [cond-mat.mes-hall]].
  
  \bibitem{LL-textbook}
  L.~D.~Landau and E.~M.~Lifshitz, ``Quantum Mechanics,'' Pergamon Press.
  
  
  \bibitem{Marino:2012zq} 
  M.~Mari\~no,
  ``Lectures on non-perturbative effects in large $N$ gauge theories, matrix models and strings,''
  Fortsch.\ Phys.\  {\bf 62}, 455 (2014)
  [arXiv:1206.6272 [hep-th]].
  
\bibitem{Dorigoni:2014hea} 
  D.~Dorigoni,
  ``An Introduction to Resurgence, Trans-Series and Alien Calculus,''
  arXiv:1411.3585 [hep-th].
  
  \bibitem{Dunne:2015eaa} 
  G.~V.~Dunne and M.~\"Unsal,
  ``What is QFT? Resurgent trans-series, Lefschetz thimbles, and new exact saddles,''
  PoS LATTICE {\bf 2015}, 010 (2016)
  [arXiv:1511.05977 [hep-lat]].



\bibitem{Keio} 
  T.~Fujimori, S.~Kamata, T.~Misumi, M.~Nitta and N.~Sakai,
  ``Resurgence Structure to All Orders of Multi-bions in Deformed SUSY Quantum Mechanics,''
  PTEP {\bf 2017}, no. 8, 083B02 (2017)
  [arXiv:1705.10483 [hep-th]].

 \bibitem{ZJJ3} 
  U.~D.~Jentschura and J.~Zinn-Justin,
  ``Instantons in quantum mechanics and resurgent expansions,''
  Phys.\ Lett.\ B {\bf 596}, 138 (2004)
  [hep-ph/0405279].


  
  
  
  
\bibitem{DU0} 
  G.~V.~Dunne and M.~\"Unsal,
  ``Generating nonperturbative physics from perturbation theory,''
  Phys.\ Rev.\ D {\bf 89}, no. 4, 041701 (2014)
  [arXiv:1306.4405 [hep-th]].

\bibitem{DU1} 
  G.~V.~Dunne and M.~\"Unsal,
  ``Uniform WKB, Multi-instantons, and Resurgent Trans-Series,''
  Phys.\ Rev.\ D {\bf 89}, no. 10, 105009 (2014)
  [arXiv:1401.5202 [hep-th]].
  
  
\bibitem{DU-Mathieu} 
  G.~V.~Dunne and M.~\"Unsal,
  ``WKB and Resurgence in the Mathieu Equation,''
  arXiv:1603.04924 [math-ph].
  
\bibitem{Stark}
N.~Hoe, B.~D'Etat, J.~Grumberg, M.~Caby, E.~Leboucher and G.~Coulaud,
``Stark effect of hydrogenic ions,''
Phys. Rev. A {\bf 25} (1982) 891.
  
  \bibitem{ZJJ1} 
  J.~Zinn-Justin and U.~D.~Jentschura,
  ``Multi-instantons and exact results I: Conjectures, WKB expansions, and instanton interactions,''
  Annals Phys.\  {\bf 313}, 197 (2004)
  [quant-ph/0501136].

\bibitem{ZJJ2} 
  J.~Zinn-Justin and U.~D.~Jentschura,
  ``Multi-instantons and exact results II: Specific cases, higher-order effects, and numerical calculations,''
  Annals Phys.\  {\bf 313}, 269 (2004)
  [quant-ph/0501137].

  

\bibitem{Gahramanov:2015yxk} 
  I.~Gahramanov and K.~Tezgin,
  ``Remark on the Dunne-\"Unsal relation in exact semiclassics,''
  Phys.\ Rev.\ D {\bf 93}, no. 6, 065037 (2016)
  [arXiv:1512.08466 [hep-th]].
  
  \bibitem{BDU} 
  G.~Basar, G.~V.~Dunne and M.~Unsal,
  ``Quantum Geometry of Resurgent Perturbative/Nonperturbative Relations,''
  JHEP {\bf 1705}, 087 (2017)
  [arXiv:1701.06572 [hep-th]].
  
    \bibitem{RP}
A.~R{\"u}dinger and F. Pi{\'e}chon, ``Hofstadter rules and generalized dimensions of the spectrum of Harper's equation,''
J. Phys. A: Math. Gen. {\bf 30} (1997) 117.
  
  \bibitem{Dorigoni:2017smz} 
  D.~Dorigoni and P.~Glass,
  ``The grin of Cheshire cat resurgence from supersymmetric localization,''
  arXiv:1711.04802 [hep-th].
  
\bibitem{CSESV-1} 
  R.~Couso-Santamar\'ia, J.~D.~Edelstein, R.~Schiappa and M.~Vonk,
  ``Resurgent Transseries and the Holomorphic Anomaly: Nonperturbative Closed Strings in Local ${\mathbb{C}\mathbb{P}^2}$,''
  Commun.\ Math.\ Phys.\  {\bf 338}, no. 1, 285 (2015)
  [arXiv:1407.4821 [hep-th]].

\bibitem{CSESV-2} 
  R.~Couso-Santamar\'ia, J.~D.~Edelstein, R.~Schiappa and M.~Vonk,
  ``Resurgent Transseries and the Holomorphic Anomaly,''
  Annales Henri Poincare {\bf 17}, no. 2, 331 (2016)
  [arXiv:1308.1695 [hep-th]].

\bibitem{CSMS} 
  R.~Couso-Santamar\'ia, M.~Marino and R.~Schiappa,
  ``Resurgence Matches Quantization,''
  J.\ Phys.\ A {\bf 50}, no. 14, 145402 (2017)
  [arXiv:1610.06782 [hep-th]].
  
  \bibitem{CM} 
  S.~Codesido and M.~Marino,
  ``Holomorphic Anomaly and Quantum Mechanics,''
  arXiv:1612.07687 [hep-th].
  
  \bibitem{CMS} 
  S.~Codesido, M.~Marino and R.~Schiappa,
  ``Non-Perturbative Quantum Mechanics from Non-Perturbative Strings,''
  arXiv:1712.02603 [hep-th].
  
\bibitem{MSW} 
  M.~Marino, R.~Schiappa and M.~Weiss,
  ``Nonperturbative Effects and the Large-Order Behavior of Matrix Models and Topological Strings,''
  Commun.\ Num.\ Theor.\ Phys.\  {\bf 2}, 349 (2008)
  [arXiv:0711.1954 [hep-th]].
  
\bibitem{EZ-face}
http://wayback.cecm.sfu.ca/projects/EZFace/

\end{thebibliography}
\end{document}